\documentclass[12pt,eqno,epsf]{article}
\usepackage{amsmath,amssymb,graphicx}
\numberwithin{equation}{section}

\def\ignore#1{{}}

\newcounter{sxn}

\newcounter{axn}

\date{}

\newdimen\mybaselineskip
\renewcommand{\thefootnote}{\arabic{footnote}}

\newcommand{\beeq}{\begin{equation}}
\newcommand{\eneq}{\end{equation}}
\newcommand{\beqn}{\begin{eqnarray}}
\newcommand{\eeqn}{\end{eqnarray}}

\newcommand{\alp}{\alpha}
\newcommand{\bt}{\beta}
\newcommand{\gm}{\gamma}
\newcommand{\Gm}{\Gamma}
\newcommand{\dlt}{\delta}

\newcommand{\tht}{\theta}

\newcommand{\kp}{\kappa}
\newcommand{\lmd}{\lambda}
\newcommand{\Lmd}{\Lambda}
\newcommand{\sgm}{\sigma}

\newcommand{\vph}{\varphi}
\newcommand{\omg}{\omega}

\newcommand{\dalp}{\dot{\alpha}}

\newcommand{\be}{\begin{equation}}
\newcommand{\ee}{\end{equation}}
\newcommand{\bea}{\begin{eqnarray}}
\newcommand{\eea}{\end{eqnarray}}
\newcommand{\eql}{\!\!\!&=\!\!\!&}

\newcommand{\sma}{\!\!\!&\simeq\!\!\!&}
\newcommand{\defa}{\!\!\!&\equiv\!\!\!&}

\newcommand{\simlt}{\stackrel{<}{{}_\sim}}

\newcommand{\tl}[1]{\tilde{#1}}
\newcommand{\bdm}[1]{{\mbox{\boldmath $#1$}}}

\newcommand{\diag}{{\rm diag}}
\newcommand{\der}{\partial}
\newcommand{\dr}{\!\!d}

\newcommand{\ie}{{i.e.}}
\newcommand{\id}{\mbox{\boldmath $1$}}

\newcommand{\brkt}[1]{\left( #1 \right)}
\newcommand{\brc}[1]{\left\{ #1 \right\}}
\newcommand{\sbk}[1]{\left[ #1 \right]}
\newcommand{\abs}[1]{\left| #1 \right|}

\renewcommand{\Im}{{\rm Im}\,}

\newcommand{\cD}{{\cal D}}

\newcommand{\cF}{{\cal F}}

\newcommand{\cH}{{\cal H}}

\newcommand{\cL}{{\cal L}}

\newcommand{\cO}{{\cal O}}
\newcommand{\cP}{{\cal P}}

\newcommand{\bC}{{\mathbb C}}

\newcommand{\bM}{{\mathbb M}}

\newcommand{\hc}{{\rm h.c.}}

\begin{document}
\thispagestyle{empty}

\baselineskip=12pt


\begin{flushright}
KEK-TH-2120
\end{flushright}

\baselineskip=25pt plus 1pt minus 1pt

\vskip 1.5cm

\begin{center}
{\LARGE\bf Spinning vortex braneworld} 

\vspace{1.0cm}
\baselineskip=20pt plus 1pt minus 1pt

\normalsize

{\large\bf Yutaka Sakamura}${}^{1,2}\!${\def\thefootnote{\fnsymbol{footnote}}
\footnote[1]{E-mail address: sakamura@post.kek.jp}}

\vskip 1.0em

${}^1${\small\it KEK Theory Center, Institute of Particle and Nuclear Studies, 
KEK, \\ 1-1 Oho, Tsukuba, Ibaraki 305-0801, Japan} \\ \vspace{1mm}
${}^2${\small\it Department of Particles and Nuclear Physics, \\
SOKENDAI (The Graduate University for Advanced Studies), \\
1-1 Oho, Tsukuba, Ibaraki 305-0801, Japan}

\end{center}

\vskip 1.0cm
\baselineskip=20pt plus 1pt minus 1pt

\begin{abstract}
A spinning vortex is considered in the context of the braneworld. 
We numerically analyze the profiles of a stationary solution 
in a six-dimensional U(1) gauge theory, 
and clarify their dependence on the angular velocity 
in the field space~$\omg$. 
We find that there is an upper limit on $\omg$, and the vortex configuration 
should be parameterized by the angular momentum rather than $\omg$. 
We also discuss matter modes localized on the vortex. 
We show that the vortex spin mixes the KK masses and induces nonvanishing masses to the zero-modes. 
It also resolves the degeneracy in the KK spectrum that the static vortex had. 
\end{abstract}

\newpage

\section{Introduction}
The braneworld scenario is interesting both from the phenomenological  and the string-theoretical points of view, 
and has been extensively investigated in vast amount of papers. 
However, many of them only discuss static brane configurations. 
This is mainly because such configurations are much easier to analyze, 
and moving branes generically violate the Lorentz symmetry on the branes. 
If we focus on a local region near the earth in the present universe, it may be a good approximation. 
However, when we discuss the cosmological evolution of the universe, we should take into account 
the brane motions in the past. 

In the braneworld scenario, 
it is natural to imagine that branes were actively moving and colliding with each other in the early stage of the universe. 
Such motions become slower as the universe expands, 
and eventually the branes approach static configurations. 
However, brane motions in the past may affect the cosmological history 
since they generally lead to various symmetry breakings including the Lorentz violation. 
In addition, the brane collision process can induce inflation~\cite{Dvali:1998pa,Takamizu:2004rq}. 
Hence it is quite important to understand how such brane motions affect the four-dimensional (4D) effective theory. 

There are various kinds of brane motions, such as translation, rotation, 
collision and merger of the branes, and so on. 
In particular, when a brane is a field-theoretical soliton rather than the D-branes in string theory,
it has a finite width. 
In such a case, deformations and spin of the brane are also possible. 
It is a nontrivial and intriguing subject to investigate how such brane motions affect the evolution of 
the 4D spacetime on the brane. 
This is the motivation for our work. 

The simplest setup for the braneworld scenario is a five-dimensional (5D) theory. 
The moving branes with codimension-one in 5D are discussed in Refs.~\cite{Gani:1999pvy}-\cite{Takamizu:2007ks}, 
and it has been shown that their motions affect the evolution of our 4D spacetime significantly. 
Here we will consider the next simplest case, \ie, the codimension-two case. 
In this case, a rotation of the branes in the extra-dimensional space becomes possible.\footnote{
Rotation of the D-branes in a compact space is discussed in Refs.~\cite{Iso:2015mva}-\cite{Iso:2018cwb}.} 
Specifically, we focus on a vortex soliton in six-dimensional (6D) theories. 
In this paper, as a first step for our purpose, we study a spinning vortex~\footnote{
As another example of spinning codimension-two objects, a rotating hollow cylinder constructed by a domain wall 
is discussed in Ref.~\cite{Niedermann:2018kki} in a 6D gravitational theory. 
In this case, the spin is necessary to stabilize the configuration against collapse due to the tension of the domain wall. 
} in a 6D non-gravitational theory. 
Namely, we assume that the typical energy scale of the vortex is much smaller than the 6D Planck mass, and 
all the gravitational effects are negligible. 
The spin of the soliton may be understood as a trail of a grazing collision that the brane has experienced in the past. 
To simplify the discussion, we focus on a stationary field configuration.\footnote{
In this paper, we only consider a classical motion. 
Our solution might no longer be stationary when quantum effects are taken into account. 
}
We also discuss the localized modes of the matter fields on the vortex, and the violation of the 4D Lorentz symmetry 
that they feel.

The paper is organized as follows. 
In the next section, we briefly review the ANO vortex in the 6D Abelian-Higgs model, 
and discuss the possibility of its rotation. 
In Sec.~\ref{Spin_vortex}, we extend the model to obtain a stationary solution for a spinning vortex. 
The profiles of the vortex background are numerically calculated, and their dependence on the angular velocity 
in the field space is clarified.  
In Sec.~\ref{scalar_sector}, we introduce a scalar matter field and discuss its Kaluza-Klein (KK) mass spectrum. 
In Sec.~\ref{fermion_sector}, we introduce the matter fermions and discuss how they are expanded into 
the KK modes in the presence of the spinning vortex background. 
We also comment on the violation of the 4D Lorentz symmetry in the 4D effective theory. 
Sec.~\ref{summary} is devoted to the summary. 

\section{Case of ANO vortex}
First we consider the Abrikosov-Nielsen-Olesen (ANO) vortex~\cite{Abrikosov:1956sx,Nielsen:1973cs}. 
The theory is the 6D Abelian-Higgs model whose Lagrangian is given by
\bea
 \cL \eql -\frac{1}{4}F_{MN}F^{MN}-\cD_M\phi^*\cD^M\phi-\frac{\lmd}{2}\brkt{\abs{\phi}^2-v^2}, 
\eea
where $M,N=0,1,\cdots,5$, and 
\bea
 F_{MN} \defa \der_M A_N-\der_N A_M, \nonumber\\
 \cD_M\phi \defa \brkt{\der_M-igA_M}\phi. 
\eea
The parameters~$\lmd$ and $v$ are chosen to be positive. 
Since the scalar~$\phi$ and the gauge field~$A_M$ have mass dimension 2 in 6D, 
the dimensions of the parameters are given by
\be
 [g] = -1, \;\;\;\;\;
 [\lmd] = -2, \;\;\;\;\;
 [v] = 2. 
\ee

The equations of motion are
\bea
 &&\der_MF^{MN}-2g\Im\brkt{\cD^N\phi^*\phi} = 0, \nonumber\\
 &&\cD_M\cD^M\phi-\lmd\phi\brkt{\abs{\phi}^2-v^2} = 0.  \label{EOM:ANO}
\eea

\subsection{Static background} \label{ANO_vortex}
The ANO vortex is obtained by imposing the background ansatz, 
\be
 \phi = vf(v^{1/2}r)e^{in\tht}, \;\;\;\;\;
 A_\tht = \frac{n\alp(v^{1/2}r)}{g}, \;\;\;\;\;
 A_{M\neq \tht} = 0,  \label{ANO:static}
\ee
where $(r,\tht)$ are the polar coordinates for the extra dimensions, 
\be
 x^4 = r\cos\tht, \;\;\;\;\;
 x^5 = r\sin\tht, 
\ee
and the integer~$n$ is the vortex number. 
The Hamiltonian density for this background is given by 
\bea
 \cH \eql \frac{1}{2}\sum_{M=1}^5F_{0M}^2+\frac{1}{4}\sum_{M,N=1}^5F_{MN}^2
 +\sum_{M=0}^5\abs{\cD_M\phi}^2
 +\frac{\lmd}{2}\brkt{\abs{\phi}^2-v^2}^2 \nonumber\\
 \eql \frac{n^2v\alp^{\prime 2}}{2g^2r^2}+v^3f^{\prime 2}+\frac{n^2v^2}{r^2}(1-\alp)^2f^2+\frac{\lmd v^4}{2}\brkt{f^2-1}^2. 
 \label{cH:ANOstatic}
\eea
Thus, in order to have a finite vortex tension (\ie, 4D vacuum energy density)~$\tau_3=2\pi\int_0^\infty dr\;r\cH$, 
the dimensionless functions~$f$ and $\alp$ should satisfy
\be
 \lim_{r\to\infty}f(v^{1/2}r) = \lim_{r\to\infty}\alp(v^{1/2}r) = 1.
\ee
Besides, the regularity of the fields at the origin requires
\be
 f(0) = \alp(0) = 0. 
\ee
With these boundary conditions, we obtain the (static) vortex solution by solving the equations of motion~(\ref{EOM:ANO}).

\subsection{Background ansatz for a spinning vortex}
In order to search for a spinning vortex solution, we extend the background ansatz~(\ref{ANO:static}) as
\bea
 \phi \eql vf(v^{1/2}r)e^{in\tht+i\omg t}, \;\;\;\;\;
 A_0 = \frac{\omg\bt(v^{1/2}r)}{g}, \nonumber\\
 A_\tht \eql \frac{n\alp(v^{1/2}r)}{g}, \;\;\;\;\;
 A_{M\neq 0,\tht} = 0.  \label{ANO:stationary}
\eea
In this case, the Hamiltonian density~(\ref{cH:ANOstatic}) becomes  
\bea
 \cH \eql \frac{1}{2}\brkt{\frac{\omg^2v\bt^{\prime 2}}{g^2}+\frac{n^2v\alp^{\prime 2}}{g^2r^2}}
 +\omg^2 v^2\brkt{1-\bt}^2f^2+v^3f^{\prime 2}+\frac{n^2v^2}{r^2}\brkt{1-\alp}^2f^2 \nonumber\\
 &&+\frac{\lmd v^4}{2}\brkt{f^2-1}^2. 
\eea
Thus, from the condition that the vortex tension~$\tau_3$ should be finite and the regularity at the origin, 
the dimensionless functions~$f$, $\alp$ and $\bt$ must satisfy the following boundary conditions. 
\bea
 &&\lim_{r\to\infty}f(v^{1/2}r) = \lim_{r\to\infty}\alp(v^{1/2}r) = \lim_{r\to\infty}\bt(v^{1/2}r) = 1, \nonumber\\
 &&f(0) = \alp(0) = \bt'(0) = 0.  \label{rot_ANO_BC}
\eea
Under  the ansatz~(\ref{ANO:stationary}), the equations of motion~(\ref{EOM:ANO}) become
\bea
 &&f''(\rho)+\frac{f'(\rho)}{\rho}+\brc{\tl{\omg}^2\brkt{1-\bt(\rho)}-\frac{n^2}{\rho^2}\brkt{1-\alp(\rho)}^2
 -\tl{\lmd}\brkt{f^2(\rho)-1}}f(\rho) = 0, \nonumber\\
 &&\alp''(\rho)-\frac{\alp'(\rho)}{\rho}+2\tl{g}^2\brkt{1-\alp(\rho)}f^2(\rho) = 0, \nonumber\\
 &&\bt''(\rho)+\frac{\bt'(\rho)}{\rho}+2\tl{g}^2\brkt{1-\bt(\rho)}f^2(\rho) = 0,  \label{rot_ANO_eqs}
\eea
where $\rho\equiv v^{1/2}r$ is a dimensionless radial coordinate, and 
\be
 \tl{\omg} \equiv \frac{\omg}{v^{1/2}}, \;\;\;\;\;
 \tl{\lmd} \equiv \lmd v, \;\;\;\;\;
 \tl{g} \equiv gv^{1/2}, 
\ee
are dimensionless parameters. 

In fact, (\ref{rot_ANO_eqs}) does not have a solution that satisfies the boundary conditions in (\ref{rot_ANO_BC}). 
Let us focus on a region~$\rho\gg 1$. 
There, the second term of the equation for $\bt$ is neglected and $f(\rho)\simeq 1$. 
Thus the solution behaves as 
\be
 \bt(\rho) \simeq 1+C e^{-\sqrt{2}\tl{g}\rho}, 
\ee
where $C$ is a real constant. 
When $C>0$, we find that $\bt''(\rho)>0$ and $\bt'(\rho)<0$ for $\rho\gg 1$. 
Therefore, 
\be
 \bt''(\rho) = 2\tl{g}^2\brkt{1-\bt(\rho)}f^2(\rho)-\frac{\bt'(\rho)}{\rho}
\ee
is positive for all regions because the second term on the right-hand-side, which is positive, gets bigger
and bigger as we approach the origin while the contribution of the first term, which is negative, decreases. 
This indicates that $\bt'(\rho)$ is always negative for any value of $\rho$. 
For a similar reason, $\bt'(\rho)$ is always positive when $C<0$. 
In both cases, we cannot satisfy the boundary conditions at the origin.\footnote{
In fact, $\lim_{\rho\to 0}\bt'(\rho)=-{\rm sign}\,(C)\infty$ in these cases. } 
The only possible case is $C=0$. 
In this case, $\bt(\rho)=1$ is a solution that satisfies the boundary conditions in (\ref{rot_ANO_BC}). 
However, this solution is gauge-equivalent to the static solution in Sec.~\ref{ANO_vortex}. 
Namely, a stationary spinning vortex solution does not exist in this model. 

This fact is expected from the following reason.\footnote{
The author thanks Keisuke Ohashi for suggesting this perspective. 
}
The vacuum of this model is 
\be
 \phi = ve^{i\dlt}, \;\;\;\;\;
 A_M = 0, 
\ee
where $\dlt$ is a real constant. 
The fluctuation modes around this vacuum are as follows. 
The gauge boson gets a nonzero mass via the Higgs mechanism for the breaking of the U(1) gauge symmetry. 
The scalar field~$\phi$ is decomposed as $\phi=\brkt{\vph+v}e^{i(\dlt+\chi)}$. 
The phase part~$\chi$ is the would-be NG boson and is absorbed by the gauge boson, 
and the radial part~$\vph$ gets a mass from the potential. 
Namely, no massless modes exist in the vacuum. 
This indicates that nonzero energy is necessary 
when we move the vortex in any direction. 
So we cannot rotate the vortex without an energy cost. 
This is the reason why there is no stationary spinning vortex solution in this model. 

According to the above perspective, we need a massless mode corresponding to the fluctuation 
along the phase direction in order to have a stationary spinning vortex. 
In the next section, we will extend the model in such a way.

\section{Stationary spinning vortex} \label{Spin_vortex}
\subsection{Setup}
We extend the previous model by adding an extra charged scalar field. 
The Lagrangian is given by
\bea
 \cL \eql -\frac{1}{4}F_{MN}F^{MN}-\sum_{i=1,2}\cD_M\phi_i^*\cD^M\phi_i-U, \nonumber\\
 U \eql \sum_{i=1,2}\frac{\lmd_i}{2}\brkt{\abs{\phi_i}^2-v_i^2}^2
 +\gm\abs{\phi_1}^2\abs{\phi_2}^2+U_0,  \label{extend_model}
\eea
where $\lmd_1$, $\lmd_2$, $v_1$, $v_2$ and $\gm$ are positive constants, and 
\be
 \cD_M\phi_i = \brkt{\der_M-igA_M}\phi_i. 
\ee
The constant~$U_0$ is irrelevant to the physics if we neglect the gravity. 

The mass dimensions of the parameters are
\be
 [g] = -1, \;\;\;\;\;
 [\lmd_1] = [\lmd_2] = [\gm] = -2, \;\;\;\;\;
 [v_1] = [v_2] = 2. 
\ee

In addition to the U(1) gauge symmetry, the model has U(1) global symmetry, 
which is denoted as U(1)${}_{\rm gl}$, under the transformation that changes 
the relative phase of $\phi_1$ and $\phi_2$. 

The vacuum structure of this model is summarized in Appendix~\ref{vac_strc}. 
In the following, we will focus on the case that
\be
 \gm v_1^2 > \lmd_2v_2^2, \;\;\;\;\;
 \lmd_1v_1^4 > \lmd_2v_2^4. \label{ineq:parameters}
\ee
Then the vacuum (\ie, the global minimum of $U$) is 
\be
 \abs{\phi_1} = v_1, \;\;\;\;\;
 \phi_2 = 0.  \label{VEV}
\ee
We will set the constant~$U_0$ so that the vacuum energy is zero in the following. 
Namely, 
\be
 U_0 = -\frac{\lmd_2v_2^4}{2}. 
\ee

The equations of motion are 
\bea
 &&\der_MF^{MN}-2g\Im\brkt{\cD^N\phi_1^*\phi_1+\cD^N\phi_2^*\phi_2} = 0, \nonumber\\
 &&\cD_M\cD^M\phi_1-\lmd_1\phi_1\brkt{\abs{\phi_1}^2-v_1^2}-\gm\phi_1\abs{\phi_2}^2 = 0, \nonumber\\
 &&\cD_M\cD^M\phi_2-\lmd_2\phi_2\brkt{\abs{\phi_2}^2-v_2^2}-\gm\abs{\phi_1}^2\phi_2 = 0. 
 \label{EOM:ssv}
\eea

This theory has the (static) ANO vortex solution, 
\bea
 \phi_1 \eql v_1f(v_1^{1/2}r), \;\;\;\;\;
 \phi_2 = 0, \nonumber\\
 A_\tht \eql \frac{n\alp(v_1^{1/2}r)}{g}, \;\;\;\;\;
 A_{M\neq \tht} = 0. \label{ANOlike}
\eea

\subsection{Background ansatz for the spinning vortex}
For the purpose of finding an axially-symmetric stationary spinning vortex solution, we make the following ansatz 
for the background.\footnote{
In the following, we will normalize the dimensionful quantities except for $f_2$ by $v_1$. 
} 
\bea
 \phi_1 \eql v_1f_1(v_1^{1/2}r)e^{in\tht}, \;\;\;\;\;
 \phi_2 = v_2f_2(v_1^{1/2}r)e^{i\omg t}, \nonumber\\
 A_0 \eql \frac{\omg\bt(v_1^{1/2}r)}{g}, \;\;\;\;\;
 A_\tht = \frac{n\alp(v_1^{1/2}r)}{g}, \;\;\;\;\;
 A_{M\neq 0,\tht} = 0,  \label{spinning_ansatz}
\eea
where $f_{1,2}$, $\alp$ and $\bt$ are dimensionless real functions, the integer~$n$ is the vortex number, 
and the real constant~$\omega$ is the angular velocity. 

Then the Hamiltonian density is
\bea
 \cH \eql \frac{v_1}{2g^2}\brkt{\omg^2\bt^{\prime 2}+\frac{n^2\alp^{\prime 2}}{r^2}}
 +v_1^2\brc{\omg^2\bt^2f_1^2+v_1f_1^{\prime 2}+\frac{n^2(1-\alp)^2}{r^2}f_1^2} \nonumber\\
 &&+v_2^2\brc{\omg^2\brkt{1-\bt}^2f_2^2+v_1f_2^{\prime 2}+\frac{n^2\alp^2}{r^2}f_2^2} \nonumber\\
 &&+\frac{\lmd_1v_1^4}{2}\brkt{f_1^2-1}^2+\frac{\lmd_2v_2^4}{2}f_2^2\brkt{f_2^2-2}
 +\gm v_1^2v_2^2f_1^2f_2^2.  \label{cH:1}
\eea
In order to have a finite vortex tension, we should require the boundary conditions at infinity. 
\bea
 \lim_{r\to\infty}f_1(v_1^{1/2}r) \eql \lim_{r\to\infty}\alp(v_1^{1/2}r) = 1, \nonumber\\
 \lim_{r\to\infty}f_2(v_1^{1/2}r) \eql \lim_{r\to\infty}\bt(v_1^{1/2}r) = 0. \label{bd_cond:inf}
\eea
From the regularity at the vortex core, we obtain the boundary conditions at the origin. 
\be
 f_1(0) = \alp(0) = 0, \;\;\;\;\;
 f_2'(0) = \bt'(0) = 0.  \label{bd_cond:0}
\ee

With our ansatz, the equations of motion in (\ref{EOM:ssv}) are translated into the equations for 
the dimensionless functions as 
\bea
 f_1''+\frac{f_1'}{\rho}+\brc{\tl{\omg}^2\bt^2-\frac{n^2}{\rho^2}\brkt{1-\alp}^2
 -\tl{\lmd}_1\brkt{f_1^2-1}-\tl{\gm}\xi f_2^2}f_1 = 0, \nonumber\\
 f_2''+\frac{f_2'}{\rho}+\brc{\tl{\omg}^2\brkt{1-\bt}^2-\frac{n^2}{\rho^2}\alp^2
 -\tl{\lmd}_2\xi\brkt{f_2^2-1}-\tl{\gm}f_1^2}f_2 = 0, \nonumber\\
 \alp''-\frac{\alp'}{\rho}+2\tl{g}^2\brc{\brkt{1-\alp}f_1^2-\xi\alp f_2^2} = 0, \nonumber\\
 \bt''+\frac{\bt'}{\rho}-2\tl{g}^2\brc{\bt f_1^2-\xi\brkt{1-\bt}f_2^2} = 0, \label{eq:bg}
\eea
where
\bea
 \rho \defa v_1^{1/2}r, \;\;\;\;\;
 \tl{\lmd}_1 \equiv \lmd_1v_1, \;\;\;\;\;
 \tl{\lmd}_2 \equiv \lmd_2v_1, \;\;\;\;\;
 \tl{\gm} \equiv \gm v_1, \nonumber\\
 \tl{g} \defa gv_1^{1/2}, \;\;\;\;\;
 \tl{\omg} \equiv \frac{\omg}{v_1^{1/2}}, \;\;\;\;\;
 \xi \equiv \frac{v_2^2}{v_1^2} 
\eea
are dimensionless coordinate and parameters. 

Due to the U(1)${}_{\rm gl}$, this model has a massless mode corresponding to the fluctuation 
changing the relative phase between $\phi_1$ and $\phi_2$ at every spacetime point. 
Thus it is expected for the above equations to have a solution that satisfies 
the boundary conditions~(\ref{bd_cond:inf}) and (\ref{bd_cond:0}), 
in contrast to the previous model.

\subsection{Asymptotic behaviors of the solution}
From the equations in (\ref{eq:bg}) with the boundary conditions~(\ref{bd_cond:inf}) and (\ref{bd_cond:0}), 
we can read off the asymptotic behaviors of the dimensionless functions. 
In a region~$\rho\ll 1$, they behave as 
\bea
 f_1(\rho) \eql C_{f_1}^0\rho^{\abs{n}}\brc{1+\cO(\rho^2)}, \nonumber\\
 f_2(\rho) \eql C_{f_2}^0+\cO(\rho^2), \nonumber\\
 \alp(\rho) \eql C_\alp^0\rho^2+\cO(\rho^4), \nonumber\\
 \bt(\rho) \eql C_\bt^0+\cO(\rho^2),  \label{bg:asymp:core}
\eea
where $C_{f_1}^0$, $C_{f_2}^0$, $C_\alp^0$, and $C_\bt^0$ are real constants. 

Next we consider a region~$\rho\gg 1$. 
Then, using (\ref{bd_cond:inf}), (\ref{eq:bg}) is reduced to 
\bea
 \hat{f}_1''+\frac{\hat{f}_1'}{\rho}-\brkt{\tl{\omg}^2\bt^2-\frac{n^2}{\rho^2}\hat{\alp}^2
 +2\tl{\lmd}_1\hat{f}_1-\tl{\gm}\xi f_2^2} \sma 0, \nonumber\\
 f_2''+\frac{f_2'}{\rho}+\brkt{\tl{\omg}^2-\frac{n^2}{\rho^2}+\tl{\lmd}_2\xi-\tl{\gm}}f_2 \sma 0, \nonumber\\
 \hat{\alp}''-\frac{\hat{\alp}'}{\rho}-2\tl{g}^2\brkt{\hat{\alp}-\xi f_2^2} \sma 0, \nonumber\\
 \bt''+\frac{\bt'}{\rho}-2\tl{g}^2\brkt{\bt-\xi f_2^2} \sma 0,  \label{eq:bg:ap}
\eea
where
\be
 \hat{f}_1(\rho) \equiv 1-f_1(\rho), \;\;\;\;\;
 \hat{\alp}(\rho) \equiv 1-\alp(\rho). 
\ee
The solution of the second equation is expressed by the (modified) Bessel function as
\be
 f_2(\rho) \simeq \begin{cases} K_n(\sqrt{a_2}\rho) & (\tl{\omg}^2 < \tl{\gm}-\tl{\lmd}_2\xi) \\
 J_n(\sqrt{\abs{a_2}}\rho), \: Y_n(\sqrt{\abs{a_2}}\rho) & (\tl{\omg}^2 > \tl{\gm}-\tl{\lmd}_2\xi) \end{cases}, 
\ee
up to the normalization factor, where
\be
 a_2 \equiv \tl{\gm}-\tl{\lmd}_2\xi-\tl{\omg}^2.  \label{def:a_2}
\ee
Namely, when $a_2 > 0$, it behaves as
\be
 f_2(\rho) \simeq \frac{C_{f_2}^\infty}{\sqrt{\rho}}e^{-\sqrt{a_2}\rho},  \label{asymp:f_2}
\ee
where $C_{f_2}^\infty$ is a positive constant. 
Using this and the last two equations in (\ref{eq:bg:ap}), we find that 
\bea
 \alp(\rho) \sma \begin{cases} 1-C_\alp^\infty\sqrt{\rho}e^{-\sqrt{2}\tl{g}\rho} & (\tl{g}^2 < 2a_2) \\ 
 \displaystyle\rule{0pt}{22pt} 1-\xi\brkt{C_{f_2}^\infty}^2\frac{e^{-2\sqrt{a_2}\rho}}{\rho} & (\tl{g}^2 > 2a_2) \end{cases}, 
 \nonumber\\
 \bt(\rho) \sma \begin{cases} \displaystyle \frac{C_\bt^\infty}{\sqrt{\rho}}e^{-\sqrt{2}\tl{g}\rho}
 & (\tl{g}^2 < 2a_2) \\ \displaystyle\rule{0pt}{22pt}
 \xi \brkt{C_{f_2}^\infty}^2\frac{e^{-2\sqrt{a_2}\rho}}{\rho} & (\tl{g}^2 > 2a_2) \end{cases},  \label{profile:ab}
\eea 
where $C_\alp^\infty$ and $C_\bt^\infty$ are real constants. 
Then, from the first equation in (\ref{eq:bg:ap}) with the above asymptotic forms, we obtain 
\be
 f_1(\rho) \simeq \begin{cases} \displaystyle 1-\frac{C_{f_1}^\infty}{\sqrt{\rho}}e^{-\sqrt{2\tl{\lmd}_1}\rho} 
 & \brkt{\tl{\lmd}_1 < \min(4\tl{g}^2,2a_2)} \\
 \displaystyle\rule{0pt}{22pt} 1-\frac{n^2(C_\alp^\infty)^2-\tl{\omg}^2(C_\bt^\infty)^2}{2\tl{\lmd}_1}\frac{e^{-2\sqrt{2}\tl{g}\rho}}{\rho}
 & \brkt{4\tl{g}^2 < \min(\tl{\lmd}_1,2a_2)} \\
 \displaystyle\rule{0pt}{22pt} 1-\frac{\tl{\gm}\xi}{2\tl{\lmd}_1}\frac{\brkt{C_{f_2}^\infty}^2}{\rho}e^{-2\sqrt{a_2}\rho} 
 & \brkt{2a_2 < \min(\tl{\lmd}_1,4\tl{g}^2)} \label{profile:f1}
 \end{cases}, 
\ee
where $C_{f_1}^\infty$ is a real constant. 

Thus, when $a_2$ is small enough, the Hamiltonian density~(\ref{cH:1}) is approximated as
\bea
 \cH \eql v_1^3\left[\frac{1}{2\tl{g}^2}\brkt{\tl{\omg}^2\bt^{\prime 2}+\frac{n^2\alp^{\prime 2}}{\rho^2}}
 +\brc{\tl{\omg}^2\bt^2f_1^2+f_1^{\prime 2}+\frac{n^2(1-\alp)^2}{\rho^2}f_1^2} \right.\nonumber\\
 &&\hspace{5mm}
 +\xi\brc{\tl{\omg}^2\brkt{1-\bt}^2f_2^2+f_2^{\prime 2}+\frac{n^2\alp^2}{\rho^2}f_2^2} \nonumber\\
 &&\hspace{5mm}\left.
 +\frac{\tl{\lmd}_1}{2}\brkt{f_1^2-1}^2+\frac{\tl{\lmd}_2\xi^2}{2}f_2^2\brkt{f_2^2-2}
 +\tl{\gm}\xi f_1^2f_2^2\right] \nonumber\\
 \sma v_1^3\brkt{\xi\tl{\omg}^2f_2^2-\tl{\lmd}_2\xi^2f_2^2
 +\tl{\gm}\xi f_2^2} \nonumber\\
 \sma v_1^3\xi\brkt{C_{f_2}^\infty}^2\brkt{\tl{\omg}^2
 -\tl{\lmd}_2\xi+\tl{\gm}}\frac{e^{-2\sqrt{a_2}\rho}}{\rho}, 
\eea
for $\rho\gg 1$. 
Therefore, the vortex tension~$\tau_3\equiv 2\pi\int_0^\infty d\rho\;\rho\cH$ diverges 
when $a_2\simeq 0$. 
This indicates that there is a maximum value of the (normalized) angular velocity~$\tl{\omg}$.\footnote{
Notice that the constant~$C_{f_2}^\infty$ depends on $\tl{\omg}$. 
Thus, the actual divergent value of $\tl{\omg}$ slightly deviates from (\ref{omg_max}). 
(See the next subsection.)
} 
\be
 \tl{\omg}_{\rm max} \simeq \sqrt{\tl{\gm}-\tl{\lmd}_2\xi}.  \label{omg_max}
\ee

\subsection{Profiles of the solution} \label{profiles:sol}
\begin{figure}[!h]
\includegraphics[width=7cm]{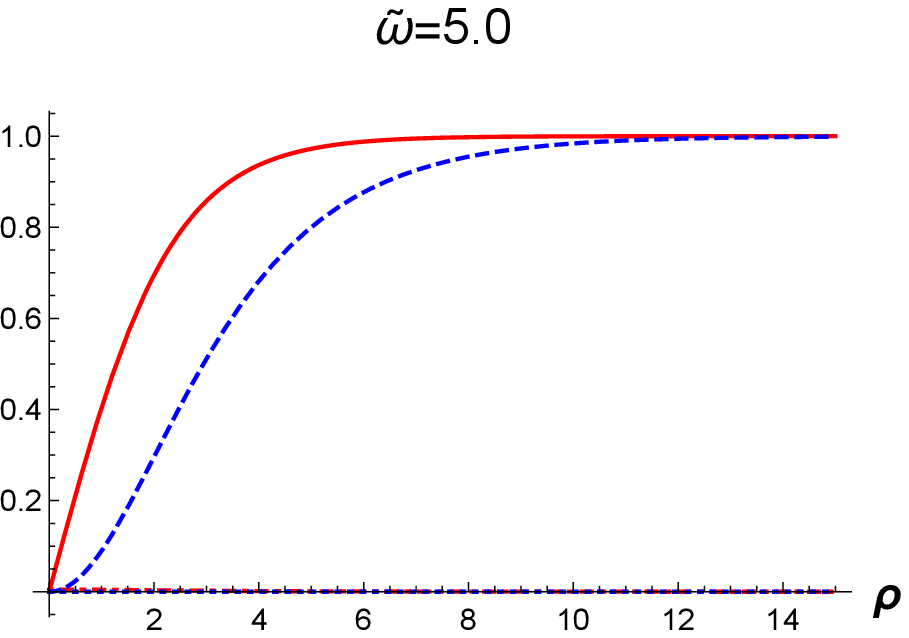}
\includegraphics[width=7cm]{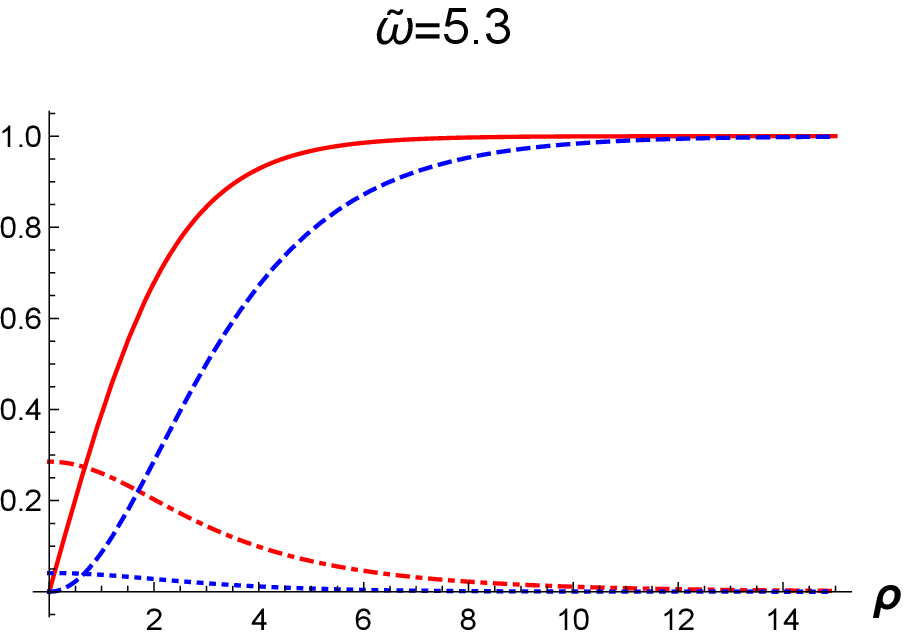}  \\
\vspace{3mm}

\includegraphics[width=7cm]{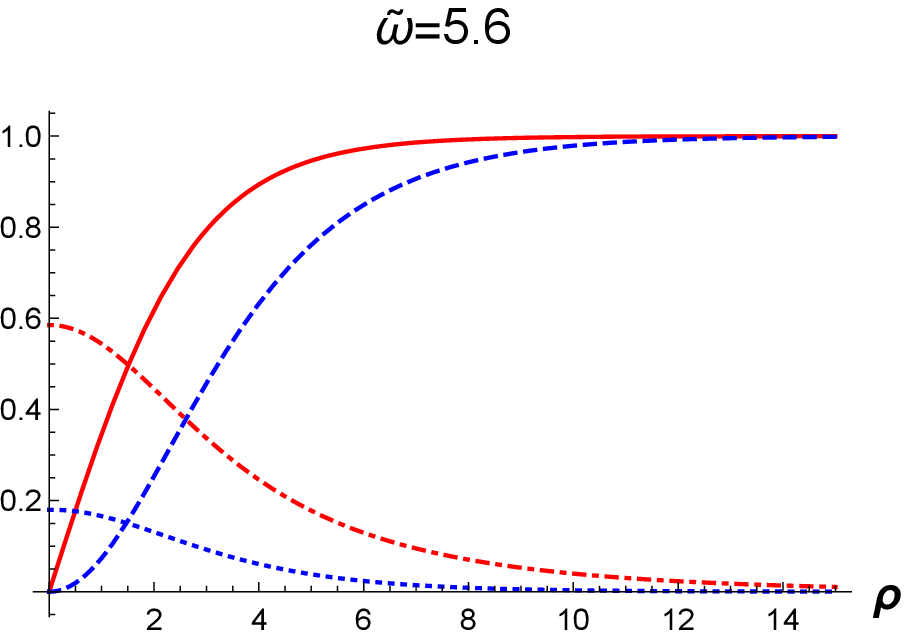}
\includegraphics[width=7cm]{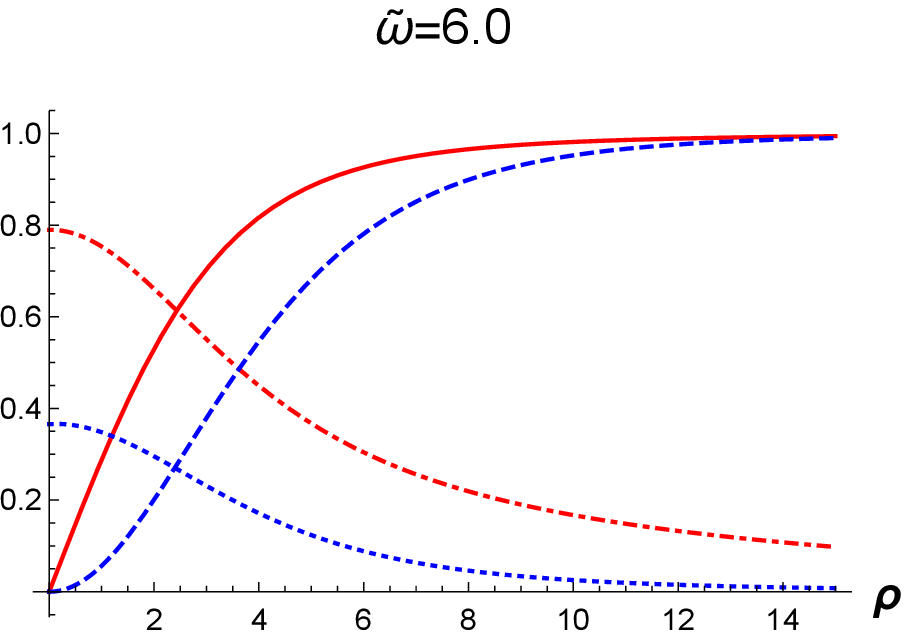} \\
\vspace{3mm}

\includegraphics[width=7cm]{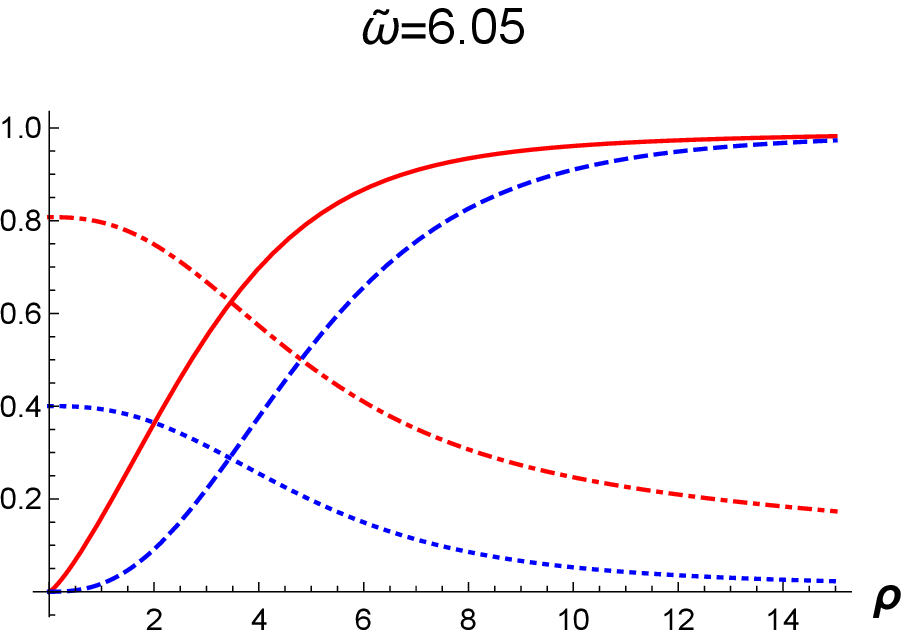}
\includegraphics[width=7cm]{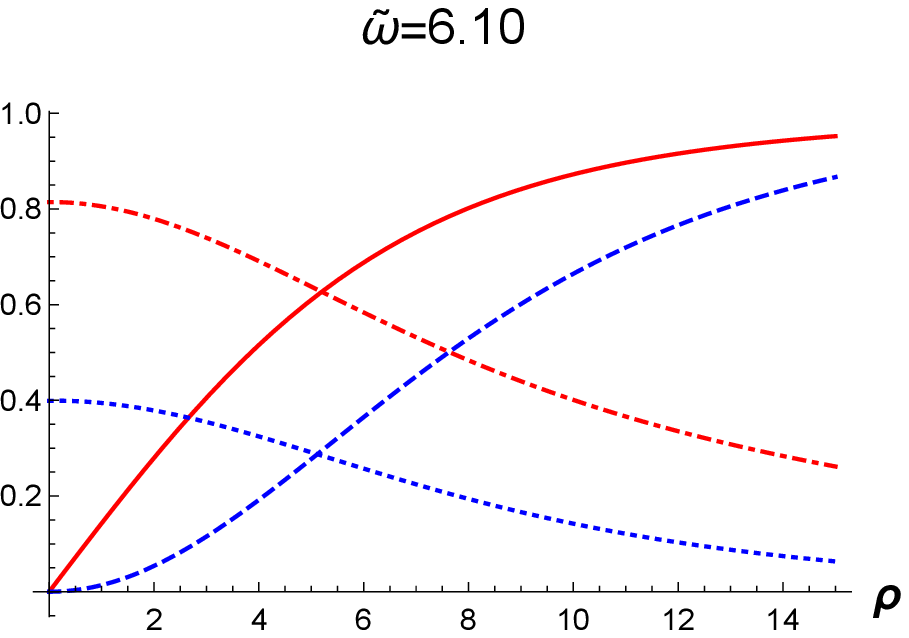}
\caption{The profiles of the dimensionless functions. 
The solid, the dot-dashed, the dashed and the dotted lines represent $f_1(\rho)$,  $f_2(\rho)$, $\alp(\rho)$ and $\bt(\rho)$, 
respectively. 
The parameters are chosen as $n=1$, $\tl{g}=0.4$, $\tl{\lmd}_1=0.3$, $\tl{\lmd}_2=0.2$, $\tl{\gm}=0.5$ and $\xi=0.7$. }
\label{profile}
\end{figure}

The equations in (\ref{eq:bg}) with the boundary conditions~(\ref{bd_cond:inf}) and (\ref{bd_cond:0}) 
can be solved numerically. 
Fig.~\ref{profile} shows the profiles of the solution. 
The solid, the dot-dashed, the dashed and the dotted lines represent $f_1(\rho)$,  $f_2(\rho)$, $\alp(\rho)$ and $\bt(\rho)$, 
respectively. 
The parameters are chosen as $n=1$, $\tl{g}=0.4$, $\tl{\lmd}_1=0.3$, $\tl{\lmd}_2=0.2$, $\tl{\gm}=0.5$ and $\xi=0.7$. 
For $\tl{\omg}\simlt 0.5$, $f_2(\rho)$ and $\bt(\rho)$ are exponentially small, and the background is 
almost that of the ANO vortex~(\ref{ANOlike}). 
For $0.5\simlt\tl{\omg}\simlt 0.61$, $f_2(\rho)$ and $\bt(\rho)$ grow as $\tl{\omg}$ increases, 
and the profiles of $f_1(\rho)$ and $\alp(\rho)$ are deformed due to the centrifugal force 
induced by the spin of the vortex. 
For $\tl{\omg}>0.61$, the functions do not decay enough in the region of $\rho\gg 1$, 
and cannot satisfy the boundary condition~(\ref{bd_cond:inf}). 

\begin{figure}[h]
\includegraphics[width=7cm]{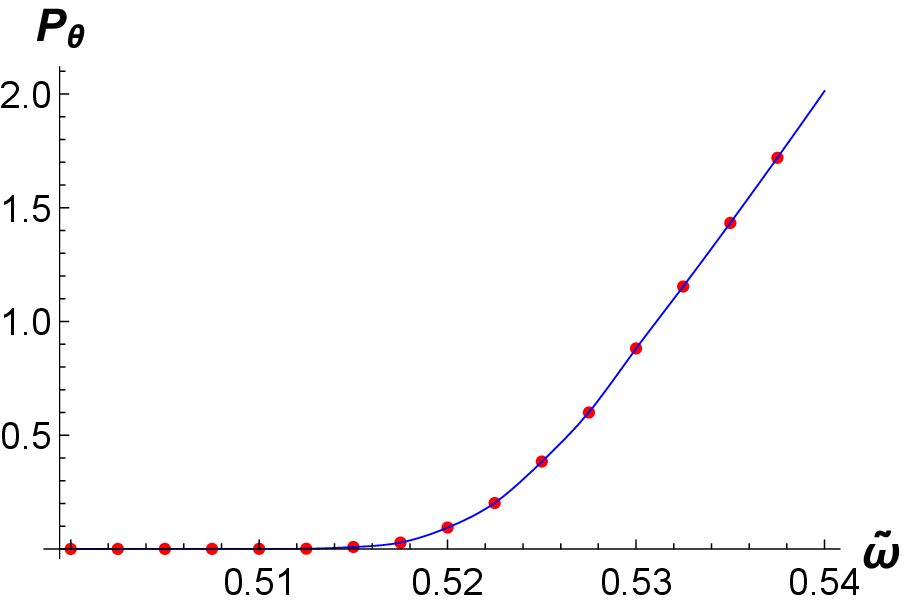} \hspace{5mm}
\includegraphics[width=7cm]{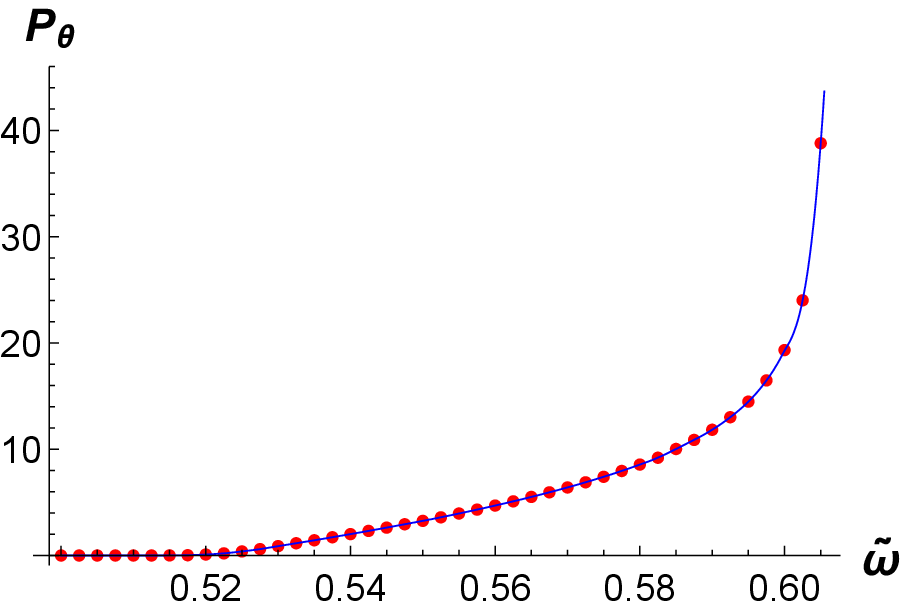}  
\caption{The angular momentum~$P_\tht$ as a function of $\tl{\omg}$. 
The parameter choice is the same as that of Fig.~\ref{profile}. 
$P_\tht$ is normalized by $v_1^{5/2}$. }
\label{Ptht-omg}
\end{figure}

These behaviors of the functions can be understood by noticing that
the centrifugal force is proportional to the angular momentum~$P_\tht$ of the vortex, rather than the angular velocity  
in the field space~$\omg$.  
The angular momentum~$P_\tht$ is given by 
\bea
 P_\tht \defa \int dx^4dx^5\;\cP_\tht = 2\pi\int_0^\infty d\rho\;\rho\cP_\tht(\rho), \nonumber\\
 \cP_\tht \defa -\brc{\cD_\tht\phi_1^*\cD^0\phi_1+\cD_\tht\phi_2^*\cD^0\phi_2+\hc}
 -F_{\tht r}F^{0r} \nonumber\\
 \eql -n\tl{\omg} v_1^{5/2}\left[\brc{1-\alp(\rho)}\bt(\rho)f_1^2(\rho)
 +\alp(\rho)\brc{1-\bt(\rho)}f_2^2(\rho)-\frac{\alp'(\rho)\bt'(\rho)}{\tl{g}^2}\right], 
 \label{def:P_tht}
\eea
where $\cP_\tht(\rho)$ is the Noether current for the rotation in the $x^4$-$x^5$ plane. 
Fig.~\ref{Ptht-omg} shows the relation between $P_\tht$ and $\omg$. 
The overlap integrals in (\ref{def:P_tht}) are exponentially small for $\tl{\omg}\simlt 0.5$, 
$P_\tht$ increases linearly with $\tl{\omg}$ in the region~$0.52\simlt\tl{\omg}\simlt 0.6$, 
and it diverges at some value around $\tl{\omg}=0.61$. 
In fact, with our parameter choice, (\ref{omg_max}) is 
\be
 \tl{\omg}_{\rm max} \simeq 0.6. 
\ee
These behaviors indicate that we should parameterize the vortex configuration by $P_\tht$ rather than by $\omg$.

\section{Localized scalar modes} \label{scalar_sector}
In this section, we introduce an additional scalar field~$\Phi$ whose U(1) charge is $q_\Phi$ as a matter field. 
Its Lagrangian is given by 
\be
 \cL_{\rm s} = -\cD^M\Phi^*\cD_M\Phi-M_\Phi^2\abs{\Phi}^2-\brkt{\kp_1\abs{\phi_1}^2+\kp_2\abs{\phi_2}^2}\abs{\Phi}^2, 
\ee
where $\kp_{1,2}>0$, and
\be
 \cD_M\Phi \equiv \brkt{\der_M-iq_\Phi gA_M}\Phi. 
\ee

The equation of motion for $\Phi$ is
\be
 \cD^M\cD_M\Phi-M_\Phi^2\Phi-\brkt{\kp_1\abs{\phi_1}^2+\kp_2\abs{\phi_2}^2}\Phi = 0. 
\ee
Substituting the background~(\ref{spinning_ansatz}), the linearized equation of motion is given by
\bea
 &&\der^\mu\der_\mu\Phi+2iq_\Phi\omg\bt(v_1^{1/2}r)\der_0\Phi \nonumber\\
 &&+\brc{\der_r^2+\frac{1}{r}\der_r+\frac{1}{r^2}\der_\tht^2-\frac{2iq_\Phi n\alp(v_1^{1/2}r)}{r^2}\der_\tht
 +q_\Phi^2\omg^2\bt^2(v_1^{1/2}r)-\frac{q_\Phi^2n^2\alp^2(v_1^{1/2}r)}{r^2}}\Phi \nonumber\\
 &&-\brc{M_\Phi^2+\kp_1v_1^2f_1^2(v_1^{1/2}r)+\kp_2v_2^2f_2^2(v_1^{1/2}r)}\Phi = 0.  \label{lin_EOM:scalar}
\eea

\subsection{Mode expansion}
The 6D scalar field~$\Phi$ is decomposed into the KK modes as
\be
 \Phi(x^\mu,r,\tht) = \sum_Kh_\Phi^{(K)}(\rho,\tht)\vph^{(K)}(x^\mu), \label{KKexpand:scalar}
\ee
where $\rho=v_1^{1/2}r$. 
We choose the mode functions as solutions of the following mode equation. 
\bea
 &&\bigg\{\der_\rho^2+\frac{1}{\rho}\der_\rho+\frac{1}{\rho^2}\der_\tht^2-\frac{2iq_\Phi n\alp(\rho)}{\rho^2}\der_\tht^2
 +q_\Phi^2\tl{\omg}^2\bt^2(\rho)-\frac{q_\Phi^2n^2\alp^2(\rho)}{\rho^2} \nonumber\\
 &&\hspace{5mm}
 -\tl{M}_\Phi^2-\tl{\kp}_1f_1^2(\rho)-\tl{\kp}_2\xi f_2^2(\rho)\bigg\}h_\Phi^{(K)}
 = -\tl{m}_K^2h_\Phi^{(K)},  \label{md_eq:scalar}
\eea
where
\be
 \tl{M}_\Phi^2 \equiv \frac{M_\Phi^2}{v_1}, \;\;\;\;\;
 \tl{m}_K^2 \equiv \frac{m_K^2}{v_1}, \;\;\;\;\;
 \tl{\kp}_1 \equiv \kp_1v_1, \;\;\;\;\;
 \tl{\kp}_2 \equiv \kp_2v_1, 
\ee
are dimensionless ($m_K$ is the KK mass). 

Since (\ref{md_eq:scalar}) has rotational symmetry in the extra dimensions, 
the eigenvalues~$\tl{m}_K^2$ in (\ref{md_eq:scalar}) have degeneracy 
and the mode functions~$h_\Phi^{(K)}(\rho,\tht)$ can be expressed as
\be
 h_\Phi^{(k)[m]}(\rho,\tht) = b_\Phi^{(k)[m]}(\rho)e^{im\tht}, \label{vrbl_separate}
\ee
where $m$ is an integer and labels the degenerate modes. 
Then, the mode equation becomes
\bea
 \brc{-\der_\rho^2-\frac{1}{\rho}\der_\rho+V(\rho)}b_\Phi^{(k)[m]}(\rho) = \tl{m}_k^2b_\Phi^{(k)[m]}(\rho), 
 \label{md_eq:b:scalar}
\eea
where
\be
 V(\rho) \equiv \frac{\brc{q_\Phi n\alp(\rho)-m}^2}{\rho^2}-q_\Phi^2\tl{\omg}^2\bt^2(\rho)+\tl{M}_\Phi^2
 +\tl{\kp}_1f_1^2(\rho)+\tl{\kp}_2\xi f_2^2(\rho).  \label{def:V_eff}
\ee
A more explicit derivation of (\ref{vrbl_separate}) and (\ref{md_eq:b:scalar}) is given in Appendix~\ref{variable_separation}. 
This has the form of a 1-dimensional Schr\"odinger equation with the potential~$V(\rho)$.

\subsection{KK spectrum}
We can obtain the KK mass spectrum by solving (\ref{md_eq:b:scalar}). 
However, it depends on many parameters and it is hard to express it analytically. 
Thus, we illustrate its property by analyzing the Schr\"odinger equation 
with the potential approximated by a simple function. 

\begin{figure}[t]
\begin{center}
\includegraphics[width=145mm]{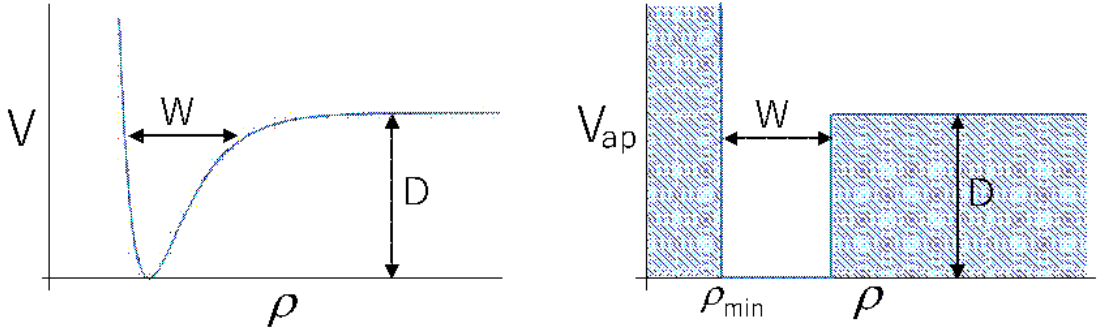} 
\end{center}
\caption{The typical form of the potential~$V(\rho)$ in the static vortex case (left figure), 
which is approximated by a simple function (right figure). }
\label{V_pot:figure}
\end{figure}

\begin{figure}[t]
\begin{center}
\includegraphics[width=6.5cm]{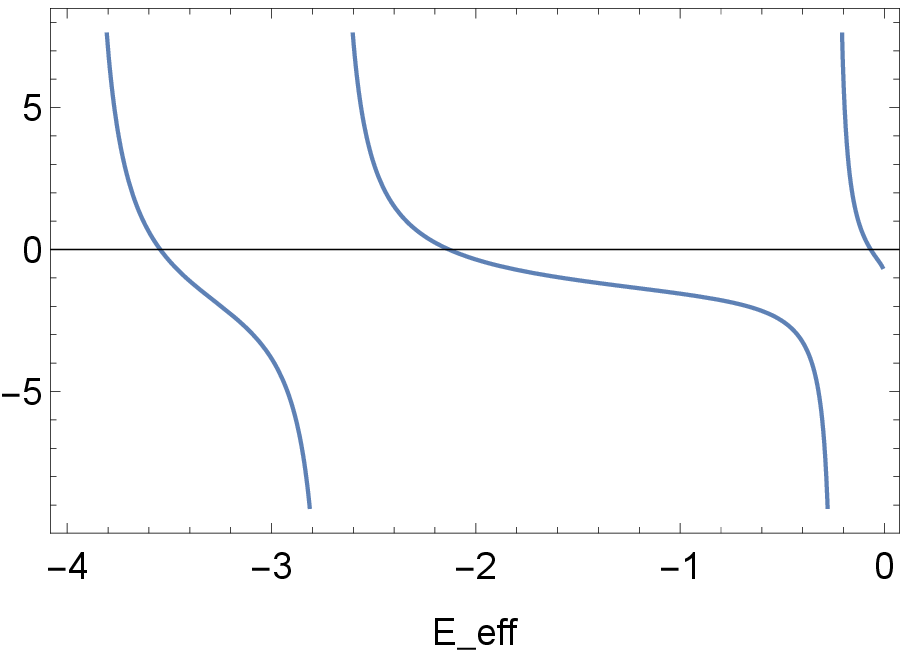}\hspace{4mm}
\includegraphics[width=6.5cm]{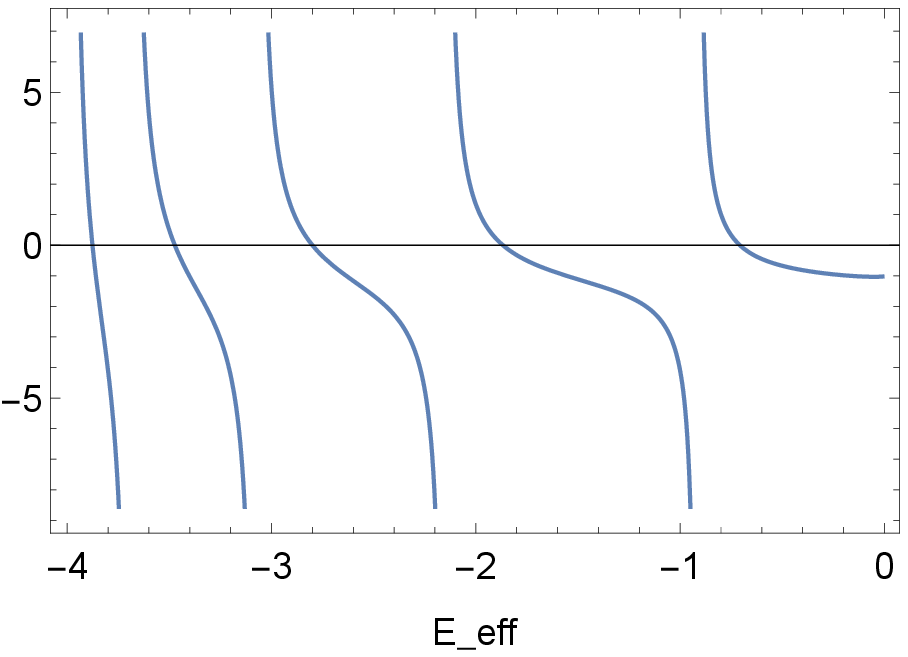} \\
\vspace{5mm} 

\includegraphics[width=6.5cm]{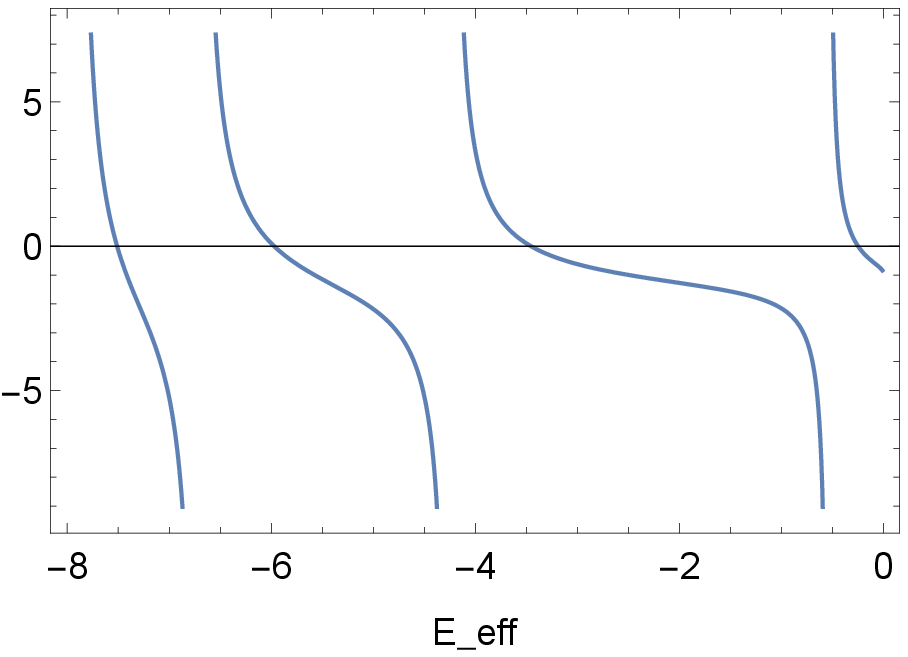}\hspace{4mm}
\includegraphics[width=6.5cm]{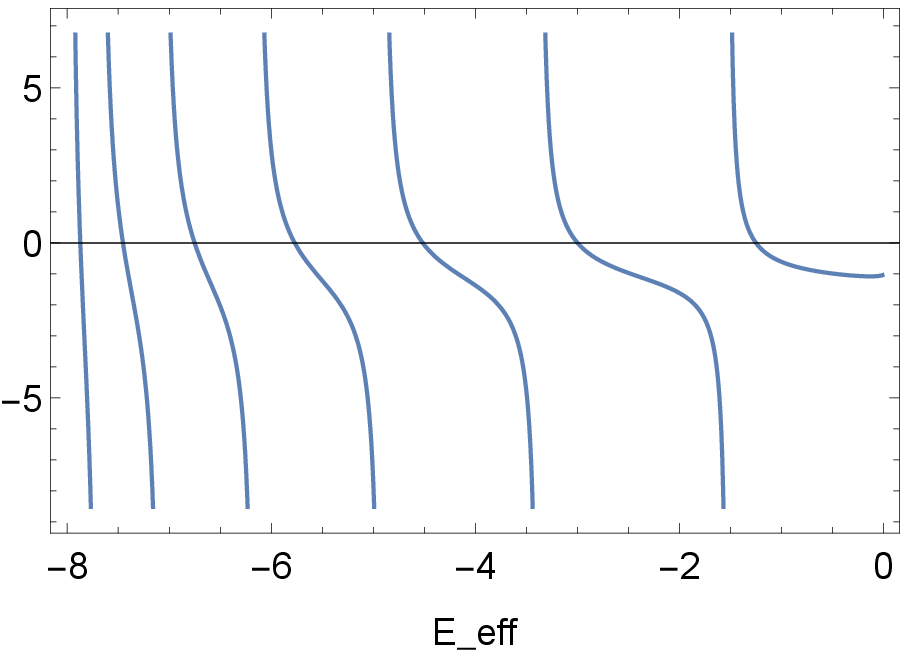}  
\end{center}
\caption{The plots of $\cF(E_{\rm eff})$ in (\ref{def:cF}) for $(W,D)=(4,4)$ (top-left), $(W,D)=(8,4)$ (top-right), 
$(W,D)=(4,8)$ (bottom left) and $(W,D)=(8,8)$ (bottom-right). 
We chose $\rho_{\rm min}$ to 1.0 for all plots.  }
\label{spctrm}
\end{figure}

Let us first consider the static vortex case (\ie, $\omg=0$). 
The typical form of the potential~$V(\rho)$ in this case is shown 
by the left figure in Fig.~\ref{V_pot:figure}.  
In order to see the properties of the spectrum in this system, we approximate $V(\rho)$ 
by the following simple function~$V_{\rm ap}(\rho)$ up to a constant. 
\be
 V_{\rm ap}(\rho) = \begin{cases} \infty & (\rho\leq \rho_{\rm min}) \\
 -D & (\rho_{\rm min}\leq \rho\leq \rho_{\rm min}+W) \\
 0 & (\rho>\rho_{\rm min}+W) \end{cases}, 
\ee
where the constants~$D$ and $W$ denote the depth and width of the potential 
(see the right figure in Fig.~\ref{V_pot:figure}). 
Thus, (\ref{md_eq:b:scalar}) is approximated by
\be
 \brc{-\der_\rho^2-\frac{1}{\rho}\der_\rho+V_{\rm ap}(\rho)}b_\Phi^{(k)[m]}(\rho) = E_{\rm eff}b_\Phi^{(k)[m]}(\rho). 
 \label{Sch_eq}
\ee
where $E_{\rm eff}\equiv \tl{m}_k^2+C_E$ ($C_E$: constant). 
\ignore{
The boundary conditions are given by
\bea
 b_\Phi^{(k)[m]}(\rho_{\rm min}) \eql 0, \nonumber\\
 b_\Phi^{(k)[m]}(\rho) \eql o(\rho^{-1}) \;\;\;\;\; (\mbox{for $\rho\gg 1$}). 
\eea
The second condition comes from the normalization condition of the mode function,\footnote{
We have taken the normalization, 
\be
 \int d\rho d\tht\;\rho \abs{h_\Phi^{(k)[m]}(\rho,\tht)}^2 = 1. 
\ee
} 
\be
 \int_0^\infty d\rho\;\rho b_\Phi^{(k)[m]2}(\rho) = \frac{1}{2\pi}. 
\ee
}
We will concentrate on the bound state solutions whose eigenvalues~$E_{\rm eff}$ satisfy $-D<E_{\rm eff}<0$. 
Then, the solution of (\ref{Sch_eq}) is
\be
 b_\Phi^{(k)[m]}(\rho) = N_{<}\brc{J_0\brkt{\sqrt{E_{\rm eff}+D}\rho}
 -\frac{J_0\brkt{\sqrt{E_{\rm eff}+D}\rho_{\rm min}}}{Y_0\brkt{\sqrt{E_{\rm eff}+D}\rho_{\rm min}}}
 Y_0\brkt{\sqrt{E_{\rm eff}+D}\rho}}, 
\ee
for $\rho_{\rm min}\leq \rho \leq \rho_{\rm min}+W$, and 
\be
 b_\Phi^{(k)[m]}(\rho) = N_{>}K_0\brkt{\sqrt{-E_{\rm eff}}\rho}, 
\ee
for $\rho>\rho_{\rm min}+W$. 
Here, $J_0(z)$ and $Y_0(z)$ are the Bessel functions of the first and second kinds, 
and $K_0(z)$ is the modified Bessel function of the second kind. 
The mode function~$b_\Phi^{(k)[m]}(\rho)$ and its derivative should be continuous at $\rho=\rho_{\rm min}+W$. 
In order for these conditions to satisfy with nonvanishing $N_{<}$ and $N_{>}$, it must be satisfied that
\be
 0 = \sqrt{-E_{\rm eff}}F_1(E_{\rm eff})K_1\brkt{\sqrt{-E_{\rm eff}}\rho_{\rm b}}
 -\sqrt{E_{\rm eff}+D}F_2(E_{\rm eff})K_0\brkt{\sqrt{-E_{\rm eff}}\rho_{\rm b}}, 
\ee
where $\rho_{\rm b}\equiv \rho_{\rm min}+W$, and 
\bea
 F_1(E_{\rm eff}) \defa J_0\brkt{\sqrt{E_{\rm eff}+D}\rho_{\rm b}}Y_0\brkt{\sqrt{E_{\rm eff}+D}\rho_{\rm min}} \nonumber\\
 &&-J_0\brkt{\sqrt{E_{\rm eff}+D}\rho_{\rm min}}Y_0\brkt{\sqrt{E_{\rm eff}+D}\rho_{\rm b}}, \nonumber\\
 F_2(E_{\rm eff}) \defa J_1\brkt{\sqrt{E_{\rm eff}+D}\rho_{\rm b}}Y_0\brkt{\sqrt{E_{\rm eff}+D}\rho_{\rm min}} \nonumber\\
 &&-J_0\brkt{\sqrt{E_{\rm eff}+D}\rho_{\rm min}}Y_1\brkt{\sqrt{E_{\rm eff}+D}\rho_{\rm b}}. 
\eea
Fig.~\ref{spctrm} shows the plots of 
\be
 \cF(E_{\rm eff}) \equiv \frac{\sqrt{-E_{\rm eff}}F_1(E_{\rm eff})K_1\brkt{\sqrt{-E_{\rm eff}}\rho_{\rm b}}}
 {\sqrt{E_{\rm eff}+D}F_2(E_{\rm eff})K_0\brkt{\sqrt{-E_{\rm eff}}\rho_{\rm b}}}-1. 
 \label{def:cF}
\ee 
The points of $\cF(E_{\rm eff})=0$ denote the eigenvalues. 
The plots show the following properties. 
\begin{itemize}
 \item The deeper the potential well of $V_{\rm ap}(\rho)$ is, the more modes are bounded to the potential. 
 \item The wider the well is, the more densely the eigenvalues are distributed. 
\end{itemize}
Since the eigenvalue of (\ref{md_eq:b:scalar}) is the square of the (normalized) KK mass, 
it must be non-negative. 
In particular, we consider a case that the lowest mass eigenvalue is zero~$\tl{m}_0^2=0$, 
which can be achieved by tuning the bulk squared mass~$M_\Phi^2$ appropriately.\footnote{
Some symmetry, such as supersymmetry, can ensure this parameter tuning. 
} 
This corresponds to the choice that the constant~$C_E$ is chosen as the lowest eigenvalue~$E_0$. 

We can numerically read off the depth~$D$ and the width~$W$ of the potential~$V(\rho)$ 
(see Fig.~\ref{V_pot:figure}). 
Qualitatively, $D$ is a decreasing function of the penetration length of the vortex, which is read off from (\ref{profile:ab}), 
and an increasing function of the correlation length, which is read off from (\ref{profile:f1}).  
The width~$W$ is a decreasing function of the penetration length while having a nontrivial dependence 
on the correlation length. 

The spin of the vortex induces nonvanishing $\bt(\rho)$ and $f_2(\rho)$. 
Both of them have support near the core of the vortex, and affect the shape of the potential~$V(\rho)$. 
However, since the signs of their contributions to the potential in (\ref{def:V_eff}) are opposite, 
their effects on $\rho_{\rm min}$, $D$ and $W$ depend on the parameters~$q_\Phi$, $\tl{\omg}$, 
$\tl{\kp}_2$ and $\xi$. 
In particular, the negative contribution~$q_\Phi^2\tl{\omg}^2\bt^2(\rho)$ in (\ref{def:V_eff}) indicates that 
there can be a negative $\tl{m}_k^2$ solution, which indicates that the background configuration is unstable. 
This reflects the fact that the vortex configuration becomes unstable when $\tl{\omg}$ exceeds 
the critical value~$\tl{\omg}_{\rm max}$ mentioned in the previous section. 

For $E_{\rm eff}>0$, (\ref{Sch_eq}) has a continuous spectrum, which corresponds to unbounded states. 
Thus, (\ref{KKexpand:scalar}) should be understood as
\be
 \Phi(x^\mu,r,\tht) = \sum_m\brc{\sum_{k=0}^{k_{\rm max}-1}h_\Phi^{(k)[m]}(\rho,\tht)\vph^{(k)[m]}(x^\mu)
 +\int_{\lmd_{\rm min}}^\infty d\lmd\;h_\Phi^{(\lmd)[m]}(\rho)\vph^{(\lmd)[m]}(x^\mu)}, 
 \label{genuine:KKexpand}
\ee
where $k_{\rm max}$ denotes the number of localized modes, and $\lmd_{\rm min}\equiv D-E_0$ 
is the lowest value of the continuous KK spectrum.

\subsection{Dispersion relations}
Making use of the orthonormal relations of the mode functions, 
we can rewrite the linearized equation of motion~(\ref{lin_EOM:scalar}) as
\be
 \der^\mu\der_\mu\vph^{(k)[m]}+2iC_{k,l}^{[m]}\der_0\vph^{(l)[m]}-m_k^2\vph^{(k)[m]} = 0, 
\ee
where
\bea
 C_{k,l}^{[m]} \defa \int d\rho d\tht\;\rho q_\Phi\omg\bt(\rho)h_\Phi^{(k)[m]*}(\rho,\tht)h_\Phi^{(l)[m]}(\rho,\tht) \nonumber\\
 \eql 2\pi\int_0^\infty d\rho\;\rho q_\Phi\omg\bt(\rho)b_\Phi^{(k)[m]}(\rho)b_\Phi^{(l)[m]}(\rho). 
\eea
If we move to the momentum basis by the Fourier transformation, this is rewritten as
\be
 \brc{\brkt{E^2-\vec{p}^2-m_k^2}\dlt_{k,l}-2C_{k,l}^{[m]}E}\tl{\vph}^{(l)[m]}(p^\mu) = 0.  \label{Ep:expression}
\ee
By diagonalizing the matrix on the left-hand-side, we obtain the dispersion relation for each KK mode. 
When the angular momentum of the vortex is small, each element of $C_{k,l}^{[m]}$ is small. 
Then, the contribution from the off-diagonal elements of $C_{k,l}^{[m]}$ to the eigenvalues of 
the matrix in (\ref{Ep:expression}) is negligible, 
and the dispersion relation for $\vph^{(k)[m]}$ is read off as
\be
 E^2-\vec{p}^2-m_k^2-2C_{k,k}^{[m]}E \simeq 0. 
\ee
Thus, the energy is expressed as~\footnote{
The other solution~$E\simeq C_{k,k}^{[m]}-\sqrt{\vec{p}^2+m_k^2+C_{k,k}^{[m]2}}$ corresponds to the annihilation mode 
of the anti-particle. 
We should also note that the energies of the particle and the anti-particle are not degenerate 
since the Lorentz symmetry is violated in our case. 
\label{otherE}
}
\bea
 E \sma C_{k,k}^{[m]}+\sqrt{\vec{p}^2+m_k^2+C_{k,k}^{[m]2}} \nonumber\\
 \eql C_{k,k}^{[m]}+\sqrt{m_k^2+C_{k,k}^{[m]2}}+\frac{\vec{p}^2}{2\sqrt{m_k^2+C_{k,k}^{[m]2}}}+\cO(\vec{p}^4). 
\eea
Therefore, we can identify the effective KK masses~$M_k$ as 
\be
 M_k \simeq \sqrt{m_k^2+C_{k,k}^{[m]2}}. 
\ee
Even if a massless localized mode exists in the static vortex case, 
it will obtain a nonvanishing mass by the spin of the vortex. 
Furthermore, we should note that this effective mass depends on the KK label~$m$. 
This means that the degeneracy in the KK spectrum, which (\ref{md_eq:scalar}) has, is resolved by the spin.

\section{Localized fermion modes} \label{fermion_sector}
In this section, we introduce matter fermions in the bulk, 
and consider the localized modes on the vortex brane. 
We introduce 6D Weyl fermions~$\Psi_\pm$ whose Lagrangian is given by
\be
 \cL_{\rm f} = \sum_{\chi_6=\pm}i\bar{\Psi}_{\chi_6}\Gm^M\cD_M\Psi_{\chi_6}
 +\brc{\brkt{y_1\phi_1+y_2\phi_2}\bar{\Psi}_-\Psi_++\hc}, 
\ee
where $\chi_6$ denotes the 6D chirality, and 
\be
 \cD_M\Psi_{\pm} = \brkt{\der_M-iq_\pm gA_M}\Psi_{\pm}. 
\ee
The constants~$q_\pm$ are the U(1) charges of $\Psi_\pm$, respectively. 
Due to the charge conservation, they are related as
\be
 q_+-q_-+1 = 0.  \label{rel:charges}
\ee
The coupling constants~$y_1$ and $y_2$ are chosen to be real and have the mass dimension 
\be
 [y_1] = [y_2] = -1. 
\ee
The notations for the gamma matrices and the fermions are collected in Appendix~\ref{notations}. 

\subsection{Mode equations}
The equations of motion for the fermions are 
\bea
 i\Gm^M\cD_M\Psi_++\brkt{y_1\bar{\phi}_1+y_2\bar{\phi}_2}\Psi_- \eql 0, \nonumber\\
 i\Gm^M\cD_M\Psi_-+\brkt{y_1\phi_1+y_2\phi_2}\Psi_+ \eql 0. 
\eea
In the 2-component spinor notation, these are rewritten as
\bea
 i\sgm^\mu\brkt{\der_\mu-iq_+gA_\mu}\bar{\zeta}_+-\brkt{\der_4+i\der_5}\chi_+
 +iq_+g\brkt{A_4+iA_5}\chi_+\brkt{y_1\bar{\phi}_1+y_2\bar{\phi}_2}\chi_- \eql 0, \nonumber\\
 i\bar{\sgm}^\mu\brkt{\der_\mu-iq_+gA_\mu}\chi_++\brkt{\der_4-i\der_5}\bar{\zeta}_+
 -iq_+g\brkt{A_4-iA_5}\bar{\zeta}_++\brkt{y_1\bar{\phi}_1+y_2\bar{\phi}_2}\bar{\zeta}_- \eql 0, \nonumber\\
 i\sgm^\mu\brkt{\der_\mu-igq_-A_\mu}\bar{\zeta}_--\brkt{\der_4-i\der_5}\chi_-
 +iq_-g\brkt{A_4-iA_5}\chi_-+\brkt{y_1\phi_1+y_2\phi_2}\chi_+ \eql 0, \nonumber\\
 i\bar{\sgm}^\mu\brkt{\der_\mu-iq_-gA_\mu}\chi_-+\brkt{\der_4+i\der_5}\bar{\zeta}_-
 -iq_-g\brkt{A_4+iA_5}\bar{\zeta}_-+\brkt{y_1\phi_1+y_2\phi_2}\bar{\zeta}_+ \eql 0, \nonumber\\
\eea
where the 2-component spinors~$\chi_\pm$ and $\bar{\zeta}_\pm$ are defined in Appendix~\ref{notations}. 

Since the background~(\ref{spinning_ansatz}) breaks the 4D Lorentz symmetry~SO(1,3) to SO(3), 
we need not discriminate the dotted and undotted indices. 
Thus, the linearized equations of motion for the fermions are expressed as 
\bea
 &&-\brkt{i\der_0+q_+\omg\bt}\bar{\zeta}_++i\sgm^i\der_i\bar{\zeta}_+ \nonumber\\
 &&-e^{i\tht}\brc{\der_r+\frac{i}{r}\brkt{\der_\tht-iq_+n\alp}}\chi_+
 +\brkt{y_1v_1f_1e^{-in\tht}+y_2v_2f_2e^{-i\omg t}}\chi_- = 0, \nonumber\\
 &&-\brkt{i\der_0+q_+\omg\bt}\chi_+-i\sgm^i\der_i\chi_+ \nonumber\\
 &&+e^{-i\tht}\brc{\der_r-\frac{i}{r}\brkt{\der_\tht-iq_+n\alp}}\bar{\zeta}_+
 +\brkt{y_1v_1f_1e^{-in\tht}+y_2v_2f_2e^{-i\omg t}}\bar{\zeta}_- = 0, \nonumber\\
 &&-\brkt{i\der_0+q_-\omg\bt}\bar{\zeta}_-+i\sgm^i\der_i\bar{\zeta}_- \nonumber\\
 &&-e^{-i\tht}\brc{\der_r-\frac{i}{r}\brkt{\der_\tht-iq_-n\alp}}\chi_-
 +\brkt{y_1v_1f_1e^{in\tht}+y_2v_2f_2e^{i\omg t}}\chi_+ = 0, \nonumber\\
 &&-\brkt{i\der_0+q_-\omg\bt}\chi_--i\sgm^i\der_i\chi_- \nonumber\\
 &&+e^{i\tht}\brc{\der_r+\frac{i}{r}\brkt{\der_\tht-iq_-n\alp}}\bar{\zeta}_-
 +\brkt{y_1v_1f_1e^{in\tht}+y_2v_2f_2e^{i\omg t}}\bar{\zeta}_+ = 0.  \label{lin_eq:fermion}
\eea
We have used the polar coordinates for the extra dimensions. 

Each component of the fermions is decomposed into the KK modes as
\bea
 \chi_\pm(x^\mu,r,\tht) \eql \sum_K h_{\rm R\pm}^{(K)}(\rho,\tht)\eta^{(K)}(x^\mu), \nonumber\\
 \bar{\zeta}_\pm(x^\mu,r,\tht) \eql \sum_K h_{\rm L\pm}^{(K)}(\rho,\tht)\eta^{(K)}(x^\mu), 
\eea
where $\rho=v_1^{1/2}r$ is the dimensionless coordinate.\footnote{
Since we do not discriminate the undotted and dotted indices, each KK mode can reside in all fermions. }
The sums in the above expansion contain integrals over the continuous spectrum, 
just like in (\ref{genuine:KKexpand}). 
We choose the mode functions as solutions of the following mode equations. 
\bea
 -e^{i\tht}\brc{\der_\rho+\frac{q_+n\alp(\rho)}{\rho}+\frac{i}{\rho}\der_\tht}h_{\rm R+}^{(K)}
 +\brc{\tl{y}_1f_1(\rho)e^{-in\tht}+\tl{y}_2\xi^{1/2}f_2(\rho)e^{-i\omg t}}h_{\rm R-}^{(K)}
 \eql \tl{m}_Kh_{\rm L+}^{(K)}, \nonumber\\
 e^{-i\tht}\brc{\der_\rho-\frac{q_+n\alp(\rho)}{\rho}-\frac{i}{\rho}\der_\tht}h_{\rm L+}^{(K)}
 +\brc{\tl{y}_1f_1(\rho)e^{-in\tht}+\tl{y}_2\xi^{1/2}f_2(\rho)e^{-i\omg t}}h_{\rm L-}^{(K)}
 \eql \tl{m}_Kh_{\rm R+}^{(K)}, \nonumber\\
 -e^{-i\tht}\brc{\der_\rho-\frac{q_-n\alp(\rho)}{\rho}-\frac{i}{\rho}\der_\tht}h_{\rm R-}^{(K)}
 +\brc{\tl{y}_1f_1(\rho)e^{in\tht}+\tl{y}_2\xi^{1/2}f_2(\rho)e^{i\omg t}}h_{\rm R+}^{(K)}
 \eql \tl{m}_Kh_{\rm L-}^{(K)}, \nonumber\\
 e^{i\tht}\brc{\der_\rho+\frac{q_-n\alp(\rho)}{\rho}+\frac{i}{\rho}\der_\tht}h_{\rm L-}^{(K)}
 +\brc{\tl{y}_1f_1(\rho)e^{in\tht}+\tl{y}_2\xi^{1/2}f_2(\rho)e^{i\omg t}}h_{\rm L+}^{(K)}
 \eql \tl{m}_Kh_{\rm R-}^{(K)},  \nonumber\\ \label{md_eq}
\eea
where
\be
 \tl{m}_K \equiv \frac{m_K}{v_1^{1/2}}, \;\;\;\;\;
 \tl{y}_1 \equiv y_1v_1^{1/2}, \;\;\;\;\;
 \tl{y}_2 \equiv y_2v_1^{1/2} 
\ee
are dimensionless ($m_K$ is the KK mass). 

Note that when $(h_{\rm R\pm}^{(K)},h_{\rm L\pm}^{(K)})$ are solutions with the eigenvalue~$\tl{m}_K$, 
the functions~$(-h_{\rm R\pm}^{(K)},h_{\rm L\pm}^{(K)})$ or $(h_{\rm R\pm}^{(K)},-h_{\rm L\pm}^{(K)})$ 
become solutions with the eigenvalue~$-\tl{m}_K$. 
Thus, we label the KK modes in such a way that 
\be
 \tl{m}_{-K} = -\tl{m}_K, \;\;\;\;\;
 h_{\rm R\pm}^{(-K)}(\rho,\tht) = -h_{\rm R\pm}^{(K)}(\rho,\tht), \;\;\;\;\;
 h_{\rm L\pm}^{(-K)}(\rho,\tht) = h_{\rm L\pm}^{(K)}(\rho,\tht), \label{rel:h}
\ee
for $n>0$ ($n$ is the integer in (\ref{spinning_ansatz})), and 
\be
 \tl{m}_{-K} = -\tl{m}_K, \;\;\;\;\;
 h_{\rm R\pm}^{(-K)}(\rho,\tht) = h_{\rm R\pm}^{(K)}(\rho,\tht), \;\;\;\;\;
 h_{\rm L\pm}^{(-K)}(\rho,\tht) = -h_{\rm L\pm}^{(K)}(\rho,\tht), \label{rel:h2}
\ee
for $n<0$. 
These are consistent with the fact that only one chiral component has zero-modes, 
which is explicitly shown in Appendix~\ref{wYukawa}.\footnote{
We have assumed that $\tl{y}_1\neq 0$ and $\tl{y}_2=0$ to specify the situation. 
} 
Making use of (\ref{md_eq}) and performing the partial integrals, we can show that
\be
 \tl{m}_K\int\dr\rho d\tht\;\rho\brkt{h_{\rm R+}^{(K)*}h_{\rm R+}^{(L)}+h_{\rm R-}^{(K)*}h_{\rm R-}^{(L)}}
 = \tl{m}_L\int\dr\rho d\tht\;\rho\brkt{h_{\rm L+}^{(K)*}h_{\rm L+}^{(L)}+h_{\rm L-}^{(K)*}h_{\rm L-}^{(L)}}, 
\ee
which leads to 
\be
 \brkt{\tl{m}_K-\tl{m}_L}\int\dr\rho d\tht\;\rho\brkt{h_{\rm R+}^{(K)*}h_{\rm R+}^{(L)}+h_{\rm R-}^{(K)*}h_{\rm R-}^{(L)}
 +h_{\rm L+}^{(K)*}h_{\rm L+}^{(L)}+h_{\rm L-}^{(K)*}h_{\rm L-}^{(L)}} = 0. 
\ee
Thus, we normalize the mode functions so that
\be
 \int\dr\rho d\tht\;\rho\brkt{h_{\rm R+}^{(K)*}h_{\rm R+}^{(L)}+h_{\rm R-}^{(K)*}h_{\rm R-}^{(L)}
 +h_{\rm L+}^{(K)*}h_{\rm L+}^{(L)}+h_{\rm L-}^{(K)*}h_{\rm L-}^{(L)}} = \dlt_{K,L}. \label{orth_norm}
\ee

\subsection{$\bdm{y_2=0}$ case}
To make the discussion more specific, we focus on the case of $y_2=0$ in the following. 
Similar to the scalar case in the previous section, the eigenvalues~$\tl{m}_K$ in (\ref{md_eq}) have degeneracy, and 
we can separate the variables as
\bea
 h_{\rm R+}^{(k)[m]}(\rho,\tht) \eql b_{\rm R+}^{(k)[m]}(\rho)e^{im\tht}, \nonumber\\
 h_{\rm R-}^{(k)[m]}(\rho,\tht) \eql b_{\rm R-}^{(k)[m]}(\rho)e^{i(m+n+1)\tht}, \nonumber\\
 h_{\rm L+}^{(k)[m]}(\rho,\tht) \eql b_{\rm L+}^{(k)[m]}(\rho)e^{i(m+1)\tht}, \nonumber\\
 h_{\rm L-}^{(k)[m]}(\rho,\tht) \eql b_{\rm L-}^{(k)[m]}(\rho)e^{i(m+n)\tht}. 
\eea
Then the mode equations in (\ref{md_eq}) are expressed as
\bea
 -\brc{\der_\rho+\frac{q_+n\alp(\rho)-m}{\rho}}b_{\rm R+}^{(k)[m]}(\rho)
 +\tl{y}_1f_1(\rho)b_{\rm R-}^{(k)[m]} \eql \tl{m}_kb_{\rm L+}^{(k)[m]}(\rho), \nonumber\\
  -\brc{\der_\rho-\frac{q_-n\alp(\rho)-m-n-1}{\rho}}b_{\rm R-}^{(k)[m]}(\rho)
 +\tl{y}_1f_1(\rho)b_{\rm R+}^{(k)[m]} \eql \tl{m}_kb_{\rm L-}^{(k)[m]}(\rho), \nonumber\\
 \brc{\der_\rho-\frac{q_+n\alp(\rho)-m-1}{\rho}}b_{\rm L+}^{(k)[m]}(\rho)
 +\tl{y}_1f_1(\rho)b_{\rm L-}^{(k)[m]}(\rho) \eql \tl{m}_kb_{\rm R+}^{(k)[m]}(\rho), \nonumber\\
 \brc{\der_\rho+\frac{q_-n\alp(\rho)-m-n}{\rho}}b_{\rm L-}^{(k)[m]}(\rho)
 +\tl{y}_1f_1(\rho)b_{\rm L+}^{(k)[m]}(\rho) \eql \tl{m}_kb_{\rm R-}^{(k)[m]}(\rho).  \label{md_eq:b}
\eea
The relations in (\ref{rel:h}) are translated into 
\be
 \tl{m}_{-k} = -\tl{m}_k, \;\;\;\;\;
 b_{\rm R\pm}^{(-k)[m]}(\rho) = -b_{\rm R\pm}^{(k)[m]}(\rho), \;\;\;\;\;
 b_{\rm L\pm}^{(-k)[m]}(\rho) = b_{\rm L\pm}^{(k)[m]}(\rho).  \label{rel:b}
\ee
Since $h_{\rm R\pm}^{(k)[m]}(\rho,\tht)$ and $h_{\rm L\pm}^{(k)[m]}(\rho,\tht)$ are normalized by (\ref{orth_norm}), 
$b_{\rm R\pm}^{(k)[m]}(\rho)$ and $b_{\rm L\pm}^{(k)[m]}(\rho)$ satisfy 
\be
 \int_0^\infty d\rho\;\rho\brkt{b_{\rm R+}^{(k)[m]}b_{\rm R+}^{(l)[m]}+b_{\rm R-}^{(k)[m]}b_{\rm R-}^{(l)[m]}
 +b_{\rm L+}^{(k)[m]}b_{\rm L+}^{(l)[m]}+b_{\rm L-}^{(k)[m]}b_{\rm L-}^{(l)[m]}} = \frac{\dlt_{k,l}}{2\pi}.  
 \label{orth_norm:2}
\ee

Making use of the orthonormal relations, we can rewrite (\ref{lin_eq:fermion}) as the linearized equations for the KK modes. 
\be
 -i\der_0\eta^{(k)[m]}+i\sum_l B_{k,l}^{[m]}\sgm^i\der_i\eta^{(l)[m]}-\sum_l C_{k,l}^{[m]}\eta^{(l)[m]}+m_k\eta^{(k)[m]} = 0, 
 \label{lin_EOM:KK}
\ee
where
\bea
 B_{k,l}^{[m]} \defa \int\dr\rho d\tht\;\rho\brkt{-h_{\rm R+}^{(k)[m]*}h_{\rm R+}^{(l)[m]}-h_{\rm R-}^{(k)[m]*}h_{\rm R-}^{(l)[m]}
 +h_{\rm L+}^{(k)[m]*}h_{\rm L+}^{(l)[m]}+h_{\rm L-}^{(k)[m]*}h_{\rm L-}^{(l)[m]}} \nonumber\\
 \eql 2\pi\int_0^\infty d\rho\;\rho\brkt{-b_{\rm R+}^{(k)[m]}b_{\rm R+}^{(l)[m]}-b_{\rm R-}^{(k)[m]}b_{\rm R-}^{(l)[m]}
 +b_{\rm L+}^{(k)[m]}b_{\rm L+}^{(l)[m]}+b_{\rm L-}^{(k)[m]}b_{\rm L-}^{(l)[m]}}, \nonumber\\
 C_{k,l}^{[m]} \defa \int\dr\rho d\tht\;\rho\omg\bt\brkt{q_+h_{\rm R+}^{(k)[m]*}h_{\rm R+}^{(l)[m]}+q_-h_{\rm R-}^{(k)[m]*}h_{\rm R-}^{(l)[m]}
 +q_+h_{\rm L+}^{(k)[m]*}h_{\rm L+}^{(l)[m]}+q_-h_{\rm L-}^{(k)[m]*}h_{\rm L-}^{(l)[m]}} \nonumber\\
 \eql 2\pi\omg\int_0^\infty d\rho\;\rho\bt\brkt{q_+b_{\rm R+}^{(k)[m]}b_{\rm R+}^{(l)[m]}
 +q_-b_{\rm R-}^{(k)[m]}b_{\rm R-}^{(l)[m]}+q_+b_{\rm L+}^{(k)[m]}b_{\rm L+}^{(l)[m]}
 +q_-b_{\rm L-}^{(k)[m]}b_{\rm L-}^{(l)[m]}}. \nonumber\\
\eea
From (\ref{rel:b}) and (\ref{orth_norm:2}), we can see that
\be
 B_{k,l}^{[m]} = \dlt_{k,-l}. 
\ee
Note also that the KK modes with different $m$ are decoupled from each other in the linearized equations of motion~(\ref{lin_EOM:KK}). 
Thus, (\ref{lin_EOM:KK}) is rewritten as
\bea
 &&i\der_0\begin{pmatrix} \bdm{\eta}_+^{[m]} \\ \eta^{(0)[m]} \\ \bdm{\eta}_-^{[m]}  \end{pmatrix}
 -i\begin{pmatrix} \bdm{0} & \bdm{0} & \id \\ \bdm{0} & 1 & \bdm{0} \\ \id & \bdm{0} & \bdm{0} \end{pmatrix}
 \sgm^i\der_i\begin{pmatrix} \bdm{\eta}_+^{[m]} \\ \eta^{(0)[m]} \\ \bdm{\eta}_-^{[m]} \end{pmatrix} \nonumber\\
 &&+\begin{pmatrix} \bC_+^{[m]} & \bdm{C}_{0+}^{[m]} & \bC_-^{[m]} \\ 
 \bdm{C}_{0+}^{[m]t} & C_{0,0}^{[m]} & \bdm{C}_{0-}^{[m]t} \\ \bC_-^{[m]} & \bdm{C}_{0-}^{[m]} & \bC_+^{[m]} \end{pmatrix}
 \begin{pmatrix} \bdm{\eta}_+^{[m]} \\ \eta^{(0)[m]} \\ \bdm{\eta}_-^{[m]}  \end{pmatrix}
 -\begin{pmatrix} \bM & \bdm{0} & \bdm{0} \\ \bdm{0} & 0 & \bdm{0} \\ 
 \bdm{0} & \bdm{0} & -\bM \end{pmatrix}
 \begin{pmatrix} \bdm{\eta}_+^{[m]} \\ \eta^{(0)[m]} \\ \bdm{\eta}_-^{[m]} \end{pmatrix} = \bdm{0}, \label{lin_EOM:KK2}
\eea
where 
\bea
 \bdm{\eta}_\pm^{[m]} \defa \brkt{\eta^{(\pm 1)[m]},\eta^{(\pm 2)[m]},
 \eta^{(\pm 3)[m]},\cdots}^t, \nonumber\\
 \bM \defa \diag\brkt{m_1,m_2,m_3,\cdots}, 
\eea
and the matrices~$\bC_\pm^{[m]}$ are defined by
\bea
 (\bC_+^{[m]})_{k,l} \defa C_{k,l}^{[m]} = C_{-k,-l}^{[m]}, \nonumber\\
 (\bC_-^{[m]})_{k,l} \defa C_{k,-l}^{[m]} = C_{-k,l}^{[m]}, 
\eea
and the column vectors~$\bdm{C}_{0\pm}^{[m]}$ are defined by 
\bea
 (\bdm{C}_{0+}^{[m]})_k \defa C_{0,k}^{[m]} = C_{k,0}^{[m]}, \nonumber\\
 (\bdm{C}_{0-}^{[m]})_k \defa C_{0,-k}^{[m]} = C_{-k,0}^{[m]}, 
\eea
for $k,l>0$. 
Note that $\bC_\pm^{[m]}$ are hermitian. 
Since (\ref{lin_EOM:KK2}) is rewritten as
\be
 i\sgm^i\der_i\begin{pmatrix} \bdm{\eta}_+^{[m]} \\ \eta^{(0)[m]} \\ \bdm{\eta}_-^{[m]} \end{pmatrix}
 = \begin{pmatrix} \bC_-^{[m]} & \bdm{C}_{0-}^{[m]} & i\der_0+\bC_+^{[m]}+\bM \\ 
 \bdm{C}_{0+}^{[m]t} & i\der_0+C_{0,0}^{[m]} & \bdm{C}_{0-}^{[m]t} \\
 i\der_0+\bC_+^{[m]}-\bM & \bdm{C}_{0+}^{[m]} & \bC_-^{[m]} \end{pmatrix}
 \begin{pmatrix} \bdm{\eta}_+^{[m]} \\ \eta^{(0)[m]} \\ \bdm{\eta}_-^{[m]} \end{pmatrix}, 
\ee
we obtain 
\bea
 &&\begin{pmatrix} \bC_-^{[m]} & \bdm{C}_{0-}^{[m]} & i\der_0+\bC_+^{[m]}+\bM \\ 
 \bdm{C}_{0+}^{[m]t} & i\der_0+C_{0,0}^{[m]} & \bdm{C}_{0-}^{[m]t} \\
 i\der_0+\bC_+^{[m]}-\bM & \bdm{C}_{0+}^{[m]} & \bC_-^{[m]} \end{pmatrix}^2
 \begin{pmatrix} \bdm{\eta}_+^{[m]} \\ \eta^{(0)[m]} \\ \bdm{\eta}_-^{[m]} \end{pmatrix} \nonumber\\
 \eql \begin{pmatrix} \bC_-^{[m]} & \bdm{C}_{0-}^{[m]} & i\der_0+\bC_+^{[m]}+\bM \\ 
 \bdm{C}_{0+}^{[m]t} & i\der_0+C_{0,0}^{[m]} & \bdm{C}_{0-}^{[m]t} \\
 i\der_0+\bC_+^{[m]}-\bM & \bdm{C}_{0+}^{[m]} & \bC_-^{[m]} \end{pmatrix}
 i\sgm^i\der_i\begin{pmatrix} \bdm{\eta}_+^{[m]} \\ \eta^{(0)[m]} \\ \bdm{\eta}_-^{[m]} \end{pmatrix} \nonumber\\
 \eql i\sgm^i\der_i\begin{pmatrix} \bC_-^{[m]} & \bdm{C}_{0-}^{[m]} & i\der_0+\bC_+^{[m]}+\bM \\ 
 \bdm{C}_{0+}^{[m]t} & i\der_0+C_{0,0}^{[m]} & \bdm{C}_{0-}^{[m]t} \\
 i\der_0+\bC_+^{[m]}-\bM & \bdm{C}_{0+}^{[m]} & \bC_-^{[m]} \end{pmatrix}
 \begin{pmatrix} \bdm{\eta}_+^{[m]} \\ \eta^{(0)[m]} \\ \bdm{\eta}_-^{[m]} \end{pmatrix} \nonumber\\
 \eql i\sgm^i\der_i\brkt{i\sgm^j\der_j}\begin{pmatrix} \bdm{\eta}_+^{[m]} \\ \eta^{(0)[m]} \\ \bdm{\eta}_-^{[m]} \end{pmatrix}
 = -\vec{\nabla}^2\begin{pmatrix} \bdm{\eta}_+^{[m]} \\ \eta^{(0)[m]} \\ \bdm{\eta}_-^{[m]} \end{pmatrix}. 
\eea
In the momentum basis, this becomes
\be
 \begin{pmatrix} \bC_-^{[m]} & \bdm{C}_{0-}^{[m]} & -E+\bC_+^{[m]}+\bM \\ 
 \bdm{C}_{0+}^{[m]t} & -E+C_{0,0}^{[m]} & \bdm{C}_{0-}^{[m]t} \\
 -E+\bC_+^{[m]}-\bM & \bdm{C}_{0+}^{[m]} & \bC_-^{[m]} \end{pmatrix}^2
 \begin{pmatrix} \bdm{\tl{\eta}}_+^{[m]} \\ \tl{\eta}^{(0)[m]} \\ \bdm{\tl{\eta}}_-^{[m]} \end{pmatrix} 
 = \vec{p}^2\begin{pmatrix} \bdm{\tl{\eta}}_+^{[m]} \\ \tl{\eta}^{(0)[m]} \\ \bdm{\tl{\eta}}_-^{[m]} \end{pmatrix}, 
\ee
or 
\be
 \begin{pmatrix} \bM_{++}^{[m]} & \bdm{M}_{+0}^{[m]} & \bM_{+-}^{[m]} \\ 
 \bdm{M}_{0+}^{[m]t} & M_{00}^{[m]} & \bdm{M}_{0-}^{[m]t} \\
 \bM_{-+}^{[m]} & \bdm{M}_{-0}^{[m]} & \bM_{--}^{[m]} \end{pmatrix}
 \begin{pmatrix} \bdm{\tl{\eta}}_+^{[m]} \\ \tl{\eta}^{(0)[m]} \\ \bdm{\tl{\eta}}_-^{[m]} \end{pmatrix}
 = \vec{p}^2\begin{pmatrix} \bdm{\tl{\eta}}_+^{[m]} \\ \tl{\eta}^{(0)[m]} \\ \bdm{\tl{\eta}}_-^{[m]} \end{pmatrix}, 
 \label{mtrx:disper}
\ee
where 
\bea
 \bM_{++}^{[m]} \defa E^2-2\bC_+^{[m]} E+\brkt{\bC_+^{[m]2}+\bdm{C}_{0-}^{[m]}\bdm{C}_{0+}^{[m]t}+\bC_-^{[m]2}}
 -\sbk{\bC_+^{[m]},\bM}-\bM^2, \nonumber\\
 \bM_{+-}^{[m]} \defa -2\bC_-^{[m]}E+\brc{\bC_+^{[m]}+\bM,\bC_-^{[m]}}+\bdm{C}_{0-}^{[m]}\bdm{C}_{0-}^{[m]t}, \nonumber\\
 \bM_{-+}^{[m]} \defa -2\bC_-^{[m]}E+\brc{\bC_+^{[m]}-\bM,\bC_-^{[m]}}+\bdm{C}_{0+}^{[m]}\bdm{C}_{0+}^{[m]t}, \nonumber\\
 \bM_{--}^{[m]} \defa E^2-2\bC_+^{[m]}E+\brkt{\bC_+^{[m]2}+\bdm{C}_{0+}^{[m]}\bdm{C}_{0-}^{[m]t}+\bC_-^{[m]2}}
 +\sbk{\bC_+^{[m]},\bM}-\bM^2, \nonumber\\
 M_{00}^{[m]} \defa E^2-2C_{0,0}^{[m]}E+C_{0,0}^{[m]2}+\bdm{C}_{0+}^{[m]t}\bdm{C}_{0-}^{[m]}
 +\bdm{C}_{0-}^{[m]t}\bdm{C}_{0+}^{[m]}, 
\eea
and
\bea
 \bdm{M}_{+0}^{[m]} \defa -\brkt{\bdm{C}_{0+}^{[m]}+\bdm{C}_{0-}^{[m]}}E+\bC_-^{[m]}\bdm{C}_{0-}^{[m]}
 +\bdm{C}_{0-}^{[m]}C_{0,0}^{[m]}+\brkt{\bC_+^{[m]}+\bM}\bdm{C}_{0+}^{[m]}, \nonumber\\
 \bdm{M}_{0+}^{[m]t} \defa -\brkt{\bdm{C}_{0+}^{[m]t}+\bdm{C}_{0-}^{[m]t}}E
 +\bdm{C}_{0+}^{[m]t}\bC_-^{[m]}+C_{0,0}^{[m]}\bdm{C}_{0+}^{[m]t}
 +\bdm{C}_{0-}^{[m]t}\brkt{\bC_+^{[m]}-\bM}, \nonumber\\
 \bdm{M}_{0-}^{[m]t} \defa -\brkt{\bdm{C}_{0+}^{[m]t}+\bdm{C}_{0-}^{[m]t}}E
 +\bdm{C}_{0-}^{[m]t}\bC_-^{[m]}+C_{0,0}^{[m]}\bdm{C}_{0-}^{[m]t}+\bdm{C}_{0+}^{[m]t}\brkt{\bC_+^{[m]}+\bM}, \nonumber\\
 \bdm{M}_{-0}^{[m]} \defa -\brkt{\bdm{C}_{0+}^{[m]}+\bdm{C}_{0-}^{[m]}}E+\bC_-^{[m]}\bdm{C}_{0+}^{[m]}
 +\bdm{C}_{0+}^{[m]}C_{0,0}^{[m]}+\brkt{\bC_+^{[m]}-\bM}\bdm{C}_{0-}^{[m]}. 
\eea
By diagonalizing the matrix on the right-hand-side of (\ref{mtrx:disper}), we can obtain 
the dispersion relations. 

Let us consider a case in which $\omg$ is sufficiently smaller than $\omg_{\rm max}$ 
in order to see how the dispersion relation for the lowest mode is distorted by the spin of the vortex. 
As we have seen in Sec.~\ref{profiles:sol}, $\bt(\rho)$ is exponentially small in such a case, and we can expand 
(\ref{mtrx:disper}) in terms of the elements of the matrix~$C_{k,l}^{[m]}$. 
We can immediately see that the contributions from the off-diagonal elements in (\ref{mtrx:disper}) 
to the eigenvalues are $\cO(\bt^2)$. 
Hence at $\cO(\bt)$, the dispersion relation for the lowest mode is read off as
\be
 E^2-2C_{0,0}^{[m]}E-\vec{p}^2 = \cO(\bt^2). 
\ee
When $\vec{p}^2\ll C_{0,0}^{[m]2}$, this is reduced to~\footnote{
The other solution~$E\simeq -\frac{\vec{p}^2}{2C_{0,0}^{[m]}}$ corresponds to the annihilation mode of 
the anti-particle. (See footnote~\ref{otherE}.)
} 
\be
 E \simeq 2C_{0,0}^{[m]}+\frac{\vec{p}^2}{2C_{0,0}^{[m]}}, 
\ee
which indicates that the lowest mode has a tiny but nonvanishing mass~$C_{0,0}^{[m]}$. 
As we mentioned in the previous section, the degeneracy in the KK spectrum is resolved 
due to the $m$-dependence of the effective masses. 

When $\omg$ is close to $\omg_{\rm max}$, we have to take into account the mixing with the higher KK modes 
by diagonalizing the matrix in (\ref{mtrx:disper}). 

In the static limit ($\omg\to 0$), the 4D Lorentz symmetry is recovered and 
(\ref{mtrx:disper}) provides ordinary relativistic dispersion relations. 
\be
 \brc{\brkt{E^2-\vec{p}^2}\id -\begin{pmatrix} \bM^2 & \bdm{0} & \bdm{0} \\
 \bdm{0} & 0 & \bdm{0} \\ \bdm{0} & \bdm{0} & -\bM^2 \end{pmatrix}}
 \begin{pmatrix} \bdm{\tl{\eta}}_+^{[m]} \\ \tl{\eta}^{(0)[m]} \\ \bdm{\tl{\eta}}_-^{[m]} \end{pmatrix}
 = \bdm{0}. 
\ee
Hence, the lowest modes become massless and chiral. 
In this limit, either $h_{\rm R\pm}^{(0)[m]}(\rho,\tht)$ or $h_{\rm L\pm}^{(0)[m]}(\rho,\tht)$ 
vanish depending on the sign of $n$, as shown in Appendix~\ref{zm_in_static}.

\ignore{
In the following analysis, we move to a gauge in which the background for $A_0$ is zero. 
Namely, the background~(\ref{spinning_ansatz}) becomes~\footnote{
In this section, the background for a field~$\Phi$ is denoted as $\Phi^{\rm bg}$
in order to discriminate it from the fluctuation mode. }
\bea
 A_\mu^{\rm bg} \eql 0, \;\;\;\;\;
 A_r^{\rm bg} = -\frac{\omg v_1^{1/2}\bt'(v_1^{1/2}r)t}{g}, \;\;\;\;\;
 A_\tht^{\rm bg} = \frac{n\alp(v_1^{1/2}r)}{g}, \nonumber\\
 \phi_1^{\rm bg} \eql v_1f_1(v_1^{1/2}r)\exp\brc{in\tht-i\omg\bt(v_1^{1/2}r)t}, \nonumber\\
 \phi_2^{\rm bg} \eql v_2f_2(v_1^{1/2}r)\exp\brc{i\omg\brkt{1-\bt(v_1^{1/2}r)}t}. 
 \label{bg:2}
\eea
Then the linearized equations of motion for the fermions are 
\bea
 i\sgm^\mu\der_\mu\bar{\zeta}_+-e^{i\tht}\brc{\brkt{\der_r-iq_+gA_r^{\rm bg}}
 +\frac{i}{r}\brkt{\der_\tht-iq_+gA_\tht^{\rm bg}}}\chi_+
 +\brkt{y_1\bar{\phi}_1^{\rm bg}+y_2\bar{\phi}_2^{\rm bg}}\chi_- \eql 0, \nonumber\\
 i\bar{\sgm}^\mu\der_\mu\chi_++e^{-i\tht}\brc{\brkt{\der_r-iq_+gA_r^{\rm bg}}
 -\frac{i}{r}\brkt{\der_\tht-iq_+gA_\tht^{\rm bg}}}\bar{\zeta}_+
 +\brkt{y_1\bar{\phi}_1^{\rm bg}+y_2\bar{\phi}_2^{\rm bg}}\bar{\zeta}_- \eql 0, \nonumber\\
 i\sgm^\mu\der_\mu\bar{\zeta}_--e^{-i\tht}\brc{\brkt{\der_r-iq_-gA_r^{\rm bg}}
 -\frac{i}{r}\brkt{\der_\tht-iq_-gA_\tht^{\rm bg}}}\chi_-
 +\brkt{y_1\phi_1^{\rm bg}+y_2\phi_2^{\rm bg}}\chi_+ \eql 0, \nonumber\\
 i\bar{\sgm}^\mu\der_\mu\chi_-+e^{i\tht}\brc{\brkt{\der_r-iq_-gA_r^{\rm bg}}
 +\frac{i}{r}\brkt{\der_\tht-iq_-gA_\tht^{\rm bg}}}\bar{\zeta}_-
 +\brkt{y_1\phi_1^{\rm bg}+y_2\phi_2^{\rm bg}}\bar{\zeta}_+ \eql 0. \nonumber\\
 \label{lin_EOM}
\eea
We have used the polar coordinates for the extra dimensions. 
Each component of the fermions is decomposed into the KK modes as
\bea
 \chi_\pm(x^\mu,r,\tht) \eql \sum_k h_{\rm R\pm}^{(k)}(\rho,\tht,\tl{t})\chi^{(k)}(x^\mu), \nonumber\\
 \bar{\zeta}_\pm(x^\mu,r,\tht) \eql \sum_k h_{\rm L\pm}^{(k)}(\rho,\tht,\tl{t})\bar{\zeta}^{(k)}(x^\mu), 
\eea
where $\rho=v_1^{1/2}r$ and $\tl{t}\equiv v_1^{1/2}t$ are dimensionless coordinates.  
The mode equations are read off from (\ref{lin_EOM}) with (\ref{bg:2}) as 
\bea
 \tl{m}_k h_{\rm L+}^{(k)} \eql -e^{i\tht}\brc{\der_\rho+\frac{q_+n\alp(\rho)}{\rho}
 +i\brkt{\frac{1}{\rho}\der_\tht+q_+\tl{\omg}\bt'(\rho)\tl{t}}}h_{\rm R+}^{(k)} \nonumber\\
 &&+\brc{\tl{y}_1f_1(\rho)e^{-in\tht+i\tl{\omg}\bt(\rho)\tl{t}}
 +\tl{y}_2f_2(\rho)e^{-i\tl{\omg}\brkt{1-\bt(\rho)}\tl{t}}}h_{\rm R-}^{(k)}, \nonumber\\
 \tl{m}_k h_{\rm R+}^{(k)} \eql e^{-i\tht}\brc{\der_\rho-\frac{q_+n\alp(\rho)}{\rho}
 +i\brkt{-\frac{1}{\rho}\der_\tht+q_+\tl{\omg}\bt'(\rho)\tl{t}}}h_{\rm L+}^{(k)} \nonumber\\
 &&+\brc{\tl{y}_1f_1(\rho)e^{-in\tht+i\tl{\omg}\bt(\rho)\tl{t}}
 +\tl{y}_2f_2(\rho)e^{-i\tl{\omg}\brkt{1-\bt(\rho)}\tl{t}}}h_{\rm L-}^{(k)}, \nonumber\\
 \tl{m}_k h_{\rm L-}^{(k)} \eql -e^{-i\tht}\brc{\der_\rho-\frac{q_-n\alp(\rho)}{\rho}
 +i\brkt{-\frac{1}{\rho}\der_\tht+q_-\tl{\omg}\bt'(\rho)\tl{t}}}h_{\rm R-}^{(k)} \nonumber\\
 &&+\brc{\tl{y}_1f_1(\rho)e^{in\tht-i\tl{\omg}\bt(\rho)\tl{t}}
 +\tl{y}_2f_2(\rho)e^{i\tl{\omg}\brkt{1-\bt(\rho)}\tl{t}}}h_{\rm R+}^{(k)}, \nonumber\\
 \tl{m}_k h_{\rm R-}^{(k)} \eql e^{i\tht}\brc{\der_\rho+\frac{q_-n\alp(\rho)}{\rho}
 +i\brkt{\frac{1}{\rho}\der_\tht+q_-\tl{\omg}\bt'(\rho)\tl{t}}}h_{\rm L-}^{(k)} \nonumber\\
 &&+\brc{\tl{y}_1f_1(\rho)e^{in\tht-i\tl{\omg}\bt(\rho)\tl{t}}
 +\tl{y}_2f_2(\rho)e^{i\tl{\omg}\brkt{1-\bt(\rho)}\tl{t}}}h_{\rm L+}^{(k)},  \label{md_eq}
\eea
where 
\be
 \tl{m}_k\equiv m_k/v_1^{1/2}, \;\;\;\;\;
 \tl{y}_1 \equiv y_1v_1^{1/2}, \;\;\;\;\;
 \tl{y}_2 \equiv y_2v_1^{1/2} 
\ee
are dimensionless ($m_k$ is the KK mass). 
In the following, we will focus on the zero-modes, \ie, $\tl{m}_0=0$. 
When both Yukawa coupling constants~$y_1$ and $y_2$ are nonvanishing, 
it is hard to separate the coordinate-dependence of the mode functions. 
So we leave the analysis in such a case to our future works, 
and consider cases in which at least one of the Yukawa couplings vanish in this paper. 
}

\ignore{
\subsection{4D Lorentz violation}
Our background~(\ref{spinning_ansatz}) breaks 4D Lorentz symmetry due to the nonvanishing $A_0^{\rm bg}$ 
and the explicit $t$-dependence of $\phi_2^{\rm bg}$. 
In the fermionic sector, this Lorentz-violating effect appears through the explicit $t$-dependence of 
the mode functions. 
When $y_2=0$, such $t$-dependences are proportional to the U(1) charges (see (\ref{y1case}) or (\ref{zero_ys_case})). 
Thus, they are cancelled due to the gauge symmetry except for the coupling to $\phi_1$. 
The situation is similar even if we also introduce matter scalar fields that do not couple to $\phi_2$. 
Namely, the 4D Lorentz violation is not observed in the matter sector at tree level in this case. 
On the other hand, when $y_1=0$ and $y_2\neq 0$, the $t$-dependences of the mode functions cannot be cancelled. 
In this case, if we move to the gauge in which 
\bea
 A_0^{\rm bg} \eql -\frac{\omg}{g}, \;\;\;\;\;
 A_i^{\rm bg} = 0 \;\;\; (i=1,2,3), \nonumber\\
 A_r^{\rm bg} \eql -\frac{\omg v_1^{1/2}\bt'(v_1^{1/2}t)t}{g}, \;\;\;\;\;
 A_\tht^{\rm bg} = \frac{n\alp(v_1^{1/2}r)}{g}, \nonumber\\
 \phi_1^{\rm bg} \eql v_1 f_1(v_1^{1/2}r)\exp\brc{in\tht-i\omg\brkt{1+\bt(v_1^{1/2}r)}t}, \nonumber\\
 \phi_2^{\rm bg} \eql v_2f_2(v_1^{1/2}r)\exp\brc{-i\omg\bt(v_1^{1/2}r)t}, 
\eea
the $t$-dependences of the mode functions become proportional to the charges, and 
are cancelled in the 4D effective theory. 
However, due to the nonvanishing $A_0^{\rm bg}$, 
the dispersion relations deviate from the relativistic ones, just like the situation in Ref.~\cite{Iso:2018sgx}. 
The fluctuations around $\phi_1^{\rm bg}$ and $\phi_2^{\rm bg}$ couple with those of $A_0^{\rm bg}$, 
$A_r^{\rm bg}$ and $A_\tht^{\rm bg}$, and directly receive the Lorentz-violating effects of the background. 
}

\section{Summary} \label{summary}
We have considered a situation that the 3-brane where we live is spinning in extra-dimensional space. 
The ANO vortex in the Abelian-Higgs model does not have a degree of freedom to rotate 
the vortex configuration without an energy cost, 
so the stationary spinning solution does not exist. 
We have extended the model by adding an extra charged scalar so that an extra U(1) global symmetry appears, 
and the stationary spinning vortex solution is allowed. 
We find that the vortex profile has a nontrivial dependence on the angular velocity in the field space~$\omg$ 
only in a limited region, and there is an upper limit on $\omg$. 
Thus the vortex configuration should be parameterized by the angular momentum for the rotation 
in the extra-dimensional space, rather than $\omg$. 
In contrast to the ANO vortex, the U(1) gauge symmetry is not restored at the core of the vortex 
due to the nonvanishing background of the second scalar~$\phi_2$. 

The spin of the vortex violates the 4D Lorentz symmetry in the effective theory. 
Due to the nonvanishing background of the temporal component of the gauge field~$A_0$, 
the dotted and the undotted spinor indices become indistinguishable. 
Hence each KK fermionic mode resides in both 4D chiral components, 
and they are described by 2-component spinors of the unbroken SO(3). 
The dispersion relations are also modified by the spin of the vortex (or the VEV of $A_0$). 
In particular, the zero-modes are mixed with higher KK modes due to the spin, 
and obtain nonvanishing masses. 
We should also note that the vortex spin resolves the degeneracy in the KK spectrum, 
which exists in the static vortex case. 

There are many directions in which we should proceed. 
We would like to generalize the situation by considering various kinds of vortices in various models, 
and extract universal properties of the spinning vortices. 
If we extend the model in a supersymmetric way, we can also discuss the SUSY-breaking effects 
in the 4D effective theory induced by the spin of a BPS vortex. 
The vortex in motion on the compact space is also an intriguing subject. 
In this paper, we have only considered classical motion. 
However, the spinning vortex may radiate some particles by a quantum effect and lose energy. 
In such a case, the vortex solution is no longer stationary, but the angular velocity will slow down, 
and the configuration will be reduced to be static. 
It would be interesting to pursue this process and study how it affects the cosmological history. 
In addition, we have neglected gravity in our analysis to simplify the discussion, 
but it is important to investigate the effects of spin on the 4D cosmological evolution 
in 6D gravitational theories. 
We will discuss these issues in subsequent papers.

\subsection*{Acknowledgements}
The author would like to thank Keisuke Ohashi and Minoru Eto for valuable comments and discussions.

\appendix

\section{Vacuum structure of the model in Sec.~\ref{Spin_vortex}} \label{vac_strc}
Here we summarize the vacuum structure of the model~(\ref{extend_model}).\footnote{
See Ref.~\cite{Davis:1988jp} for a similar setup. 
}
The minimization conditions of the potential~$U$ are 
\bea
 0 \eql \frac{\der U}{\der\phi_1^*} = \lmd_1\phi_1\brkt{\abs{\phi_1}^2-v_1^2}+\gm\phi_1\abs{\phi_2}^2, \nonumber\\
 0 \eql \frac{\der U}{\der\phi_2^*} = \lmd_2\phi_2\brkt{\abs{\phi_2}^2-v_2^2}+\gm\abs{\phi_1}^2\phi_2. 
\eea
By solving these, we find the following stationary points of $U$. 
\begin{enumerate}
\item $\phi_1=\phi_2=0$ \label{case1}

\item $\abs{\phi_1} = v_1$ and $\phi_2=0$  \label{case2}

\item $\phi_1 = 0$ and $\abs{\phi_2}=v_2$  \label{case3}

\item $\displaystyle \abs{\phi_1}^2 = \frac{\lmd_2\brkt{\lmd_1v_1^2-\gm v_2^2}}{\lmd_1\lmd_2-\gm^2}$ and 
$\displaystyle \abs{\phi_2}^2 = \frac{\lmd_1\brkt{\lmd_2v_2^2-\gm v_1^2}}{\lmd_1\lmd_2-\gm^2}$ \\
This solution is possible only when 
\bea
 &&(\lmd_1\lmd_2-\gm^2)(\lmd_1v_1^2-\gm v_2^2)>0, \nonumber\\
 &&(\lmd_1\lmd_2-\gm^2)(\lmd_2v_2^2-\gm v_1^2)>0. \label{ineq:case4}
\eea 
\label{case4}
\end{enumerate}

In order to investigate the stability of the vacua, we divide the complex scalar fields as
\be
 \phi_1 = \phi_{\rm 1R}+i\phi_{\rm 1I}, \;\;\;\;\;
 \phi_2 = \phi_{\rm 2R}+i\phi_{\rm 2I}, 
\ee
and evaluate the Hessian matrix, 
\bea
 H(U) \eql \begin{pmatrix} \displaystyle\frac{\der^2U}{\der\phi_{\rm 1R}^2} 
 & \displaystyle\frac{\der^2U}{\der\phi_{\rm 1R}\der\phi_{\rm 1I}} 
 & \displaystyle\frac{\der^2U}{\der\phi_{\rm 1R}\der\phi_{\rm 2R}} 
 & \displaystyle\frac{\der^2U}{\der\phi_{\rm 1R}\der\phi_{\rm 2I}} \\
 \rule{0pt}{23pt}\displaystyle\frac{\der^2U}{\der\phi_{\rm 1I}\der\phi_{\rm 1R}} 
 & \displaystyle\frac{\der^2U}{\der\phi_{\rm 1I}^2}
 & \displaystyle\frac{\der^2U}{\der\phi_{\rm 1I}\der\phi_{\rm 2R}} 
 & \displaystyle\frac{\der^2U}{\der\phi_{\rm 1I}\der\phi_{\rm 2I}} \\
 \rule{0pt}{23pt}\displaystyle\frac{\der^2U}{\der\phi_{\rm 2R}\der\phi_{\rm 1R}} 
 & \displaystyle\frac{\der^2U}{\der\phi_{\rm 2R}\der\phi_{\rm 1I}} 
 & \displaystyle\frac{\der^2U}{\der\phi_{\rm 2R}^2} 
 & \displaystyle\frac{\der^2U}{\der\phi_{\rm 2R}\der\phi_{\rm 2I}} \\
 \rule{0pt}{23pt}\displaystyle\frac{\der^2U}{\der\phi_{\rm 2I}\der\phi_{\rm 1R}} 
 & \displaystyle\frac{\der^2U}{\der\phi_{\rm 2I}\der\phi_{\rm 1I}} 
 & \displaystyle\frac{\der^2U}{\der\phi_{\rm 2I}\der\phi_{\rm 2R}} 
 & \displaystyle\frac{\der^2U}{\der\phi_{\rm 2I}^2} \end{pmatrix}.
\eea

For stationary point~\ref{case1}, the Hessian is 
\be
 H(U) = \diag\brkt{-2\lmd_1v_1^2,-2\lmd_1v_1^2,-2\lmd_2v_2^2,-2\lmd_2v_2^2}. 
\ee
Thus this is a local maximum. 

For stationary point~\ref{case2}, we choose $\phi_1=v_1$ without loss of generality. 
Then, we obtain 
\be
 H(U) = \diag\brkt{4\lmd_1v_1^2,0,-2\lmd_2v_2^2+2\gm v_1^2,-2\lmd_2v_2^2+2\gm v_1^2}. 
\ee
Thus, we have the NG-mode for the $\phi_{\rm 1I}$-direction, 
and the other modes are massive when $\gm v_1^2>\lmd_2 v_2^2$. 
The potential value at this vacuum is
\be
 U_{\rm pt2} = \frac{\lmd_2v_2^4}{2}+U_0. 
\ee

For stationary point~\ref{case3}, we choose $\phi_2=v_2$ without loss of generality. 
Then we obtain 
\be
 H(U) = \diag\brkt{-2\lmd_1v_1^2+2\gm v_2^2,-2\lmd_1v_1^2+2\gm v_2^2,4\lmd_2v_2^2,0}. 
\ee
Thus, we have the NG-mode for the $\phi_{\rm 2I}$-direction, 
and the other modes are massive when $\gm v_2^2>\lmd_1 v_1^2$. 
The potential value at this vacuum is
\be
 U_{\rm pt3} = \frac{\lmd_1v_1^4}{2}+U_0. 
\ee

In stationary point~\ref{case4}, we choose 
\be
 \phi_1 = \sqrt{\frac{\lmd_2(\lmd_1v_1^2-\gm v_2^2)}{\lmd_1\lmd_2-\gm^2}}, \;\;\;\;\;
 \phi_2 = \sqrt{\frac{\lmd_1(\lmd_2v_2^2-\gm v_1^2)}{\lmd_1\lmd_2-\gm^2}}, 
\ee
without loss of generality. 
Then we obtain 
\bea
 H(U) \eql \frac{4}{\lmd_1\lmd_2-\gm^2}\begin{pmatrix} \lmd_1\lmd_2\Lmd_1 
 & 0 & \gm\sqrt{\lmd_1\lmd_2\Lmd_1\Lmd_2} & 0 \\
 0 & 0 & 0 & 0 \\
 \gm\sqrt{\lmd_1\lmd_2\Lmd_1\Lmd_2} & 0 & \lmd_1\lmd_2\Lmd_2 & 0 \\
 0 & 0 & 0 & 0 \end{pmatrix}, 
\eea
where
\be
 \Lmd_1 \equiv \lmd_1v_1^2-\gm v_2^2, \;\;\;\;\;
 \Lmd_2 \equiv \lmd_2v_2^2-\gm v_1^2. 
\ee
Thus, we have two massless modes, which correspond to 
the NG-modes for the breakings of the U(1) gauge and U(1) global symmetries. 
This stationary point is a local minimum iff 
\be
 \det\brc{\frac{4}{\lmd_1\lmd_2-\gm^2}\begin{pmatrix} \lmd_1\lmd_2\Lmd_1 & \gm\sqrt{\lmd_1\lmd_2\Lmd_1\Lmd_2} \\
 \gm\sqrt{\lmd_1\lmd_2\Lmd_1\Lmd_2} & \lmd_1\lmd_2\Lmd_2 \end{pmatrix}} 
 = \frac{16\lmd_1\lmd_2\Lmd_1\Lmd_2}{\lmd_1\lmd_2-\gm^2}  \label{detH:case4}
\ee
is positive. 
Combined with the condition~(\ref{ineq:case4}), this indicates that $\Lmd_1>0$ and $\Lmd_2>0$. 
Namely, point~\ref{case2} or \ref{case3} and point~\ref{case4} cannot be local minima simultaneously. 
The potential value at this vacuum is
\be
 U_{\rm pt4} = -\frac{\gm}{2(\lmd_1\lmd_2-\gm^2)}
 \brkt{\gm\lmd_1 v_1^4-2\lmd_1\lmd_2v_1^2v_2^2+\gm\lmd_2v_2^4}+U_0. 
\ee

In the case that points~\ref{case2} and \ref{case3} are local minima, \ie, $\Lmd_2<0$ and $\Lmd_1<0$, 
we find that 
\be
 \frac{\lmd_1}{\gm} < \frac{v_2^2}{v_1^2} < \frac{\gm}{\lmd_2}, 
\ee
which indicates that 
\be
 \gm^2 > \lmd_1\lmd_2.  \label{rel:gm_lmd}
\ee
In fact, the interaction parametrized by $\gm$ prevents both scalars~$\phi_1$ and $\phi_2$ 
having nonvanishing VEVs. 

When the conditions
\be
 \gm v_1^2>\lmd_2v_2^2, \;\;\;\;\;
 \lmd_2 v_2^4 < \lmd_1 v_1^4
\ee
are satisfied, point~\ref{case2} becomes a global minimum of the potential. 

\ignore{
The differences of the potential minima are 
\bea
 U_{\rm pt2}-U_{\rm pt3} \eql \frac{\lmd_2v_2^4}{2}-\frac{\lmd_1v_1^4}{2}, \nonumber\\
 U_{\rm pt2}-U_{\rm pt4} \eql \frac{\lmd_2v_2^4}{2}
 +\frac{\gm}{2(\lmd_1\lmd_2-\gm^2)}\brkt{\gm\lmd_1 v_1^4-2\lmd_1\lmd_2 v_1^2v_2^2+\gm\lmd_2v_2^4} \nonumber\\
 \eql \frac{\lmd_1\brkt{\lmd_2v_2^2-\gm v_1^2}^2}{2(\lmd_1\lmd_2-\gm^2)} < 0, \nonumber\\
 U_{\rm pt3}-U_{\rm pt4} \eql \frac{\lmd_1v_1^4}{2}
 +\frac{\gm}{2(\lmd_1\lmd_2-\gm^2)}\brkt{\gm\lmd_1 v_1^4-2\lmd_1\lmd_2 v_1^2v_2^2+\gm\lmd_2v_2^4} \nonumber\\
 \eql \frac{\lmd_2\brkt{\lmd_1v_1^2-\gm v_2^2}^2}{2(\lmd_1\lmd_2-\gm^2)} < 0. 
\eea
Therefore, the case~\ref{case4} cannot be the global minimum. 
}

\section{Derivation of (\ref{vrbl_separate})} \label{variable_separation}
We separate the mode functions in (\ref{KKexpand:scalar}) as 
\be
 h_\Phi^{(K)}(\rho,\tht) = b_\Phi^{(K)}(\rho)c_\Phi^{(K)}(\tht). 
\ee
Substituting this into the mode function~(\ref{md_eq:scalar}), we obtain 
\bea
 \Gm^{(K)}(\rho) \defa -\frac{\rho^2}{b_\Phi^{(K)}}
 \brc{\der_\rho^2+\frac{1}{\rho}\der_\rho+\tl{\omg}^2\bt^2-\frac{n^2\alp^2}{\rho^2}-M_\Phi^2
 -\tl{\kp}_1f_1^2-\tl{\kp}_2\xi f_2^2+\tl{m}_K^2}b_\Phi^{(K)} \nonumber\\
 \eql \frac{1}{c_\Phi^{(K)}}\brkt{\der_\tht^2-2in\alp\der_\tht}c_\Phi^{(K)}. \label{def:Gm^K}
\eea
Note that $\Gm^{(K)}(\rho)$ is a function of only $\rho$. 
Since $c_\Phi^{(K)}(\tht)$ is a function of only $\tht$ and 
$\alp(\rho)$ has a nontrivial $\rho$-dependence, this equation holds only when
\bea
 \frac{\der_\tht c_\Phi^{(K)}(\tht)}{c_\Phi^{(K)}(\tht)} \eql s_1, \nonumber\\
 \frac{\der_\tht^2c_\Phi^{(K)}(\tht)}{c_\Phi^{(K)}(\tht)} \eql s_2, \nonumber\\
 \Gm^{(K)}(\rho) \eql s_2-2in\alp(\rho)s_1, \label{eq:divide}
\eea
where $s_1$ and $s_2$ are constants. 
The first two equations are solved as
\be
 c_\Phi^{(K)}(\tht) = N_c e^{s_1\tht}, \;\;\;\;\;
 s_2 = s_1^2. 
\ee
Since the mode function should be single-valued, we find that $s_1=im$, where $m$ is an integer. 
Thus, the last equation in (\ref{eq:divide}) becomes
\be
 \Gm^{(K)}(\rho)+m^2-2nm\alp(\rho) = 0. 
\ee
This is the same as (\ref{md_eq:b:scalar}). 
By choosing the normalization of $c_\Phi^{(K)}(\tht)$ as $N_c=1$, we obtain (\ref{vrbl_separate}). 

Here note that the integer~$m$ (or the constant~$s_1$) can have different values for a given value of $\tl{m}_K$ 
in (\ref{def:Gm^K}). 
This indicates that there is a degeneracy in the spectrum and $m$ labels that degeneracy. 
Hence we can write the KK label~$K$ by two integers~$k$ and $m$, where $k$ labels different KK masses.

\section{Notations } \label{notations}
Here we collect the notations for the fermions. 
For 2-component spinors, we basically follow the notations of Ref.~\cite{Wess:1992cp}. 

The 6D gamma matrices~$\Gm^M$ are chosen as 
\be
 \Gm^\mu = \begin{pmatrix} \bdm{0}_4 & \gm^\mu \\ \gm^\mu & \bdm{0}_4 \end{pmatrix}, \;\;\;\;\;
 \Gm^4 = \begin{pmatrix} \bdm{0}_4 & i\gm_5 \\ i\gm_5 & \bdm{0}_4 \end{pmatrix}, \;\;\;\;\;
 \Gm^5 = \begin{pmatrix} \bdm{0}_4 & \id_4 \\ -\id_4 & \bdm{0}_4 \end{pmatrix}, 
\ee
where $\gm_5\equiv i\gm^0\gm^1\gm^2\gm^3=\diag(\id_2,-\id_2)$. 
They satisfy
\be
 \brc{\Gm^M,\Gm^N} = -2\eta^{MN}, 
\ee
where $\eta_{MN}=\diag(-1,1,1,1,1,1)$ is the 6D Minkowski metric. 
The 4D gamma matrices~$\gm^\mu$ are decomposed as
\be
 \gm^\mu = \begin{pmatrix} \bdm{0}_2 & \sgm^\mu \\ \bar{\sgm}^\mu & \bdm{0}_2 \end{pmatrix}. 
\ee
The 6D chirality matrix~$\Gm_7$ is defined as
\be
 \Gm_7 \equiv -\Gm^0\Gm^1\cdots\Gm^5 = \begin{pmatrix} \id_4 & \bdm{0}_4 \\ \bdm{0}_4 & -\id_4 \end{pmatrix}. 
\ee

The 6D Weyl fermions~$\Psi_\pm$ are expressed by
\be
 \Psi_+ = \begin{pmatrix} \psi_+ \\ \bdm{0}_4 \end{pmatrix}, \;\;\;\;\;
 \Psi_- = \begin{pmatrix} \bdm{0}_4 \\ \psi_- \end{pmatrix}, 
\ee
where the 4-component spinors~$\psi_\pm$ are further decomposed as 
\bea
 \psi_\pm \eql \begin{pmatrix} \chi_{\pm\alp} \\ \bar{\zeta}_\pm^{\dalp} \end{pmatrix}. 
\eea

\section{Number of zero-modes} \label{zm_in_static}
Here we focus on the zero-mode solutions of the mode equations in (\ref{md_eq:b}).\footnote{
Note that these solutions are not the mass eigenstates except for the static case ($\omg=0$). 
The ``zero-modes'' in this section denote the modes with zero-eigenvalue for the mode equations in (\ref{md_eq:b}). 
} 

\subsection{In the presence of Yukawa coupling} \label{wYukawa}
The mode equations for the zero-modes~$b_{\rm R\pm}^{(0)[m]}(\rho)$ and $b_{\rm L\pm}^{(0)[m]}(\rho)$ are 
\bea
 \brc{\der_\rho+\frac{q_+n\alp(\rho)-m}{\rho}}b_{\rm R+}^{(0)[m]}-\tl{y}_1f_1(\rho)b_{\rm R-}^{(0)[m]} \eql 0, 
 \nonumber\\
 \brc{\der_\rho-\frac{q_-n\alp(\rho)-m-n-1}{\rho}}b_{\rm R-}^{(0)[m]}
 -\tl{y}_1f_1(\rho)b_{\rm R+}^{(0)[m]} \eql 0, \nonumber\\
 \brc{\der_\rho-\frac{q_+n\alp(\rho)-m-1}{\rho}}b_{\rm L+}^{(0)[m]}
 +\tl{y}_1f_1(\rho)b_{\rm L-}^{(0)[m]} \eql 0, \nonumber\\
 \brc{\der_\rho+\frac{q_-n\alp(\rho)-m-n}{\rho}}b_{\rm L-}^{(0)[m]}
 +\tl{y}_1f_1(\rho)b_{\rm L+}^{(0)[m]} \eql 0.  \label{rad:md_eq}
\eea
Note that $b_{\rm R\pm}^{(0)[m]}(\rho)$ and $b_{\rm L\pm}^{(0)[m]}(\rho)$ are decoupled in these equations. 
Thus they can be parameterized by independent labels for $m$. 
Here we will write them as $b_{\rm R\pm}^{(0)[m]}(\rho)$ and $b_{\rm L\pm}^{(0)[m']}(\rho)$ to emphasize this point. 

Let us consider the behavior for $\rho\gg 1$. 
Since $f_1(\rho),\alp(\rho)\simeq 1$ in this region and the terms proportional to $\rho^{-1}$ are neglected, 
the asymptotic forms of the mode functions are~\footnote{
The solution proportional to $e^{\abs{\tl{y}_1}\rho}$ is non-normalizable, and is excluded.}
\be
 b_{\rm R\pm}^{(0)[m]}(\rho), \: b_{\rm L\pm}^{(0)[m']}(\rho) \sim e^{-\abs{\tl{y}_1}\rho}. 
\ee
Next we consider the behavior around the vortex core~$\rho\ll 1$. 
Using (\ref{bg:asymp:core}), the solutions of (\ref{rad:md_eq}) are expressed as 
\bea
 b_{\rm R+}^{(0)[m]}(\rho) \sma A_{\rm R+}^{[m]}\rho^m+B_{\rm R+}^{[m]}\rho^{-m+\abs{n}-n}, \nonumber\\
 b_{\rm R-}^{(0)[m]}(\rho) \sma A_{\rm R-}^{[m]}\rho^{m+\abs{n}+1}+B_{\rm R-}^{[m]}\rho^{-m-n-1}, \nonumber\\
 b_{\rm L+}^{(0)[m']}(\rho) \sma A_{\rm L+}^{[m']}\rho^{-m'-1}+B_{\rm L+}^{[m']}\rho^{m'+\abs{n}+n+1}, \nonumber\\
 b_{\rm L-}^{(0)[m']}(\rho) \sma A_{\rm L-}^{[m']}\rho^{-m'+\abs{n}}+B_{\rm L-}^{[m']}\rho^{m'+n},  \label{b_asymp:origin}
\eea
where $A_{\rm R\pm}^{[m]}$, $B_{\rm R\pm}^{[m]}$, $A_{\rm L\pm}^{[m']}$ and $B_{\rm L\pm}^{[m']}$ are constants. 
The regularity at the origin requires that all the powers in (\ref{b_asymp:origin}) should be non-negative. 
This leads to the following constraints on $m$ and $m'$. 
\begin{description}
\item[Case of $\bdm{n>0}$ :]
There is no solution for $m$, and 
\be
 -n \leq m' \leq -1. 
\ee
Thus we have $n$ left-handed zero-modes. 

\item[Case of $\bdm{n<0}$ :]
There is no solution for $m'$, and
\be
 0 \leq m \leq \abs{n}-1. 
\ee
Thus we have $\abs{n}$ right-handed zero-modes. 
\end{description}
In either case, $\abs{n}$ 4D chiral fermions are obtained in the effective theory~\cite{Jackiw:1981ee}.

\ignore{
\subsection{$\bdm{y_1=0}$ and $\bdm{y_2\neq 0}$ case}
Next we consider the case that the fermions do not couple to $\phi_1$. 
Then, we can separate the variables by assuming that
\bea
 h_{\rm R+}^{(0)[m]}(\rho,\tht,\tl{t}) \eql b_{\rm R+}^{[m]}(\rho)\exp\brc{im\tht-iq_+\tl{\omg}\bt(\rho)\tl{t}}, \nonumber\\
 h_{\rm R-}^{(0)[m]}(\rho,\tht,\tl{t}) \eql b_{\rm R-}^{[m]}(\rho)\exp\brc{i(m+1)\tht+i\tl{\omg}\brkt{1-q_-\bt(\rho)}\tl{t}}, \nonumber\\
 h_{\rm L+}^{(0)[m']}(\rho,\tht,\tl{t}) \eql b_{\rm L+}^{[m']}(\rho)\exp\brc{im'\tht-iq_+\tl{\omg}\bt(\rho)\tl{t}}, \nonumber\\
 h_{\rm L-}^{(0)[m']}(\rho,\tht,\tl{t}) \eql b_{\rm L-}^{[m']}(\rho)\exp\brc{i(m'-1)\tht+i\tl{\omg}\brkt{1-q_-\bt(\rho)}\tl{t}}. 
\eea
The mode equations for $b_{\rm R\pm}^{[m]}(\rho)$ and $b_{\rm L\pm}^{[m']}(\rho)$ are 
\bea
 \brkt{\der_\rho+\frac{q_+n\alp(\rho)}{\rho}-\frac{m}{\rho}}b_{\rm R+}^{[m]}
 -\tl{y}_2f_2(\rho)b_{\rm R-}^{[m]} \eql 0, \nonumber\\
 \brkt{\der_\rho-\frac{q_-n\alp(\rho)}{\rho}+\frac{m+1}{\rho}}b_{\rm R-}^{[m]}
 -\tl{y}_2f_2(\rho)b_{\rm R+}^{[m]} \eql 0, \nonumber\\
 \brkt{\der_\rho-\frac{q_+n\alp(\rho)}{\rho}+\frac{m'}{\rho}}b_{\rm L+}^{[m']}
 +\tl{y}_2f_2(\rho)b_{\rm L-}^{[m']} \eql 0, \nonumber\\
 \brkt{\der_\rho+\frac{q_-n\alp(\rho)}{\rho}-\frac{m'-1}{\rho}}b_{\rm L-}^{[m']}
 +\tl{y}_2f_2(\rho)b_{\rm L+}^{[m']} \eql 0. \label{rad:md_eq:2}
\eea
In a region of $\rho\gg 1$, the mode functions are damped as
\bea
 b_{\rm R+}^{[m]}(\rho) \!\!\!&\sim \!\!\!& \rho^{-q_+n+m}, \;\;\;\;\;
 b_{\rm R-}^{[m]}(\rho) \sim \rho^{q_-n-m-1}, \nonumber\\
 b_{\rm L+}^{[m']}(\rho) \!\!\!&\sim \!\!\!& \rho^{q_+n-m'}, \;\;\;\;\;
 b_{\rm L-}^{[m']}(\rho) \sim \rho^{-q_-n+m'-1}. 
\eea
In contrast to the previous case, the mode functions do not decay exponentially 
because $f_2(\rho)$ in (\ref{rad:md_eq:2}) is negligible for $\rho\gg 1$. 
Since the normalization conditions require that
\be
 \int^\infty d\rho\;\rho\abs{b_{\rm R,L\pm}(\rho)}^2 < \infty, 
\ee
the integers~$m$ and $m'$ must satisfy 
\bea
 &&1+2(-q_+n+m) < -1, \;\;\;\;\;
 1+2(q_-n-m-1) < -1, \nonumber\\
 &&1+2(q_+n-m') < -1, \;\;\;\;\;
 1+2(-q_-n+m'-1) < -1. 
\eea
Namely, they are constrained as
\bea
 &&q_-n < m < q_+n-1 = (q_--1)n-1, \nonumber\\
 &&q_+n+1 = (q_--1)n+1 < m' <q_-n.  \label{cstrt:2}
\eea
We have used (\ref{rel:charges}). 
Next we consider a region~$\rho\ll 1$. 
From (\ref{rad:md_eq:2}) with (\ref{bd_cond:0}), the mode functions behave as
\bea
 b_{\rm R+}^{[m]}(\rho) \sma A_{\rm R+}^{[m]}\rho^m+B_{\rm R+}^{[m]}\rho^{-m}, \nonumber\\
 b_{\rm R-}^{[m]}(\rho) \sma A_{\rm R-}^{[m]}\rho^{m+1}+B_{\rm R-}^{[m]}\rho^{-m-1}, \nonumber\\
 b_{\rm L+}^{[m']}(\rho) \sma A_{\rm L+}^{[m']}\rho^{m'}+B_{\rm L+}^{[m']}\rho^{-m'}, \nonumber\\
 b_{\rm L-}^{[m']}(\rho) \sma A_{\rm L-}^{[m']}\rho^{m'-1}+B_{\rm L-}^{[m']}\rho^{-m'+1}. 
\eea
The regularity at the origin requires that either coefficient~$A$ or $B$ must vanish. 
For example, for a positive $m$, the coefficients~$B_{\rm R\pm}^{[m]}$ vanish 
and the mode functions are 
\be
 b_{\rm R+}^{[m]}(\rho) \simeq A_{\rm R+}^{[m]}\rho^m, \;\;\;\;\;
 b_{\rm R-}^{[m]}(\rho) \simeq A_{\rm R-}^{[m]}\rho^{m+1}. 
\ee
The coefficients~$A_{\rm R\pm}^{[m]}$ are related through the mode equations in (\ref{rad:md_eq:2}) as
\be
 2(m+1)A_{\rm R-}^{[m]}-\tl{y}_2f_2(0)A_{\rm R+}^{[m]} = 0. 
\ee
No constraints on $m$ and $m'$ come out from the regularity at the origin. 
Thus, from (\ref{cstrt:2}), we find that $(\abs{n}-2)$ right-handed zero-modes exist when $n\leq -3$, 
while $(n-2)$ left-handed zero-modes exist when $n\geq 3$. 
No zero-mode appears when $\abs{n}\leq 2$. 
}

\subsection{In the absence of Yukawa coupling}
Next, we consider the case in which the fermions do not couple to the scalar fields. 
In this case, the four equations in (\ref{md_eq}) are decoupled and can be solved independently. 
We can separate the variables by assuming that
\bea
 h_{\rm R+}^{(0)[m_+]}(\rho,\tht) \eql b_{\rm R+}^{(0)[m_+]}(\rho)e^{im_+\tht}, \nonumber\\
 h_{\rm R-}^{(0)[m_-]}(\rho,\tht) \eql b_{\rm R-}^{(0)[m_-]}(\rho)e^{im_-\tht}, \nonumber\\
 h_{\rm L+}^{(0)[m'_+]}(\rho,\tht) \eql b_{\rm L+}^{(0)[m'_+]}(\rho)e^{im'_+\tht}, \nonumber\\
 h_{\rm L-}^{(0)[m'_-]}(\rho,\tht) \eql b_{\rm L-}^{(0)[m'_-]}(\rho)e^{im'_-\tht}, 
 \label{zero_ys_case}
\eea
where $m_\pm$ and $m'_\pm$ are integers. 
The solutions are 
\bea
 b_{\rm R\pm}^{(0)[m_\pm]}(\rho) \eql C_{\rm R\pm}^{[m_\pm]}\exp\brc{\pm\int_1^\rho d\rho'\;
 \frac{-q_\pm n\alp(\rho')+m_\pm}{\rho'}}, \nonumber\\
 b_{\rm L\pm}^{(0)[m'_\pm]}(\rho) \eql C_{\rm L\pm}^{[m'_\pm]}\exp\brc{\mp\int_1^\rho d\rho'\;
 \frac{-q_\pm n\alp(\rho')+m'_\pm}{\rho'}},  \label{b_RL:0}
\eea
where $C_{\rm R\pm}^{[m_\pm]}$ and $C_{\rm L\pm}^{[m'_\pm]}$ are normalization constants. 
In a region of $\rho\gg 1$, they behave as
\bea
 b_{\rm R\pm}^{(0)[m_\pm]}(\rho) \sma C_{\rm R\pm}^{[m_\pm]}\exp\brc{\pm(-q_\pm n+m_\pm)\ln\rho}
 = C_{\rm R\pm}^{[m_\pm]}\rho^{\pm(-q_\pm n+m_\pm)}, \nonumber\\
 b_{\rm L\pm}^{(0)[m'_\pm]}(\rho) \sma C_{\rm L\pm}^{[m'_\pm]}\exp\brc{\mp(-q_\pm n+m'_\pm)\ln\rho}
 = C_{\rm L\pm}^{[m'_\pm]}\rho^{\mp(-q_\pm n+m'_\pm)}. 
\eea
The normalization conditions require that
\be
 1\pm 2(-q_\pm n+m_\pm) < -1, \;\;\;\;\;
 1\mp 2(-q_\pm n+m'_\pm) < -1.  \label{0:cstrt:1}
\ee

In a region of $\rho\ll 1$, (\ref{b_RL:0}) is approximated as
\bea
 b_{\rm R\pm}^{(0)[m_\pm]}(\rho) \sma C_{\rm R\pm}^{[m_\pm]}\exp\brkt{\pm m_\pm\ln\rho} 
 = C_{\rm R\pm}^{[m_\pm]}\rho^{\pm m_\pm}, \nonumber\\
 b_{\rm L\pm}^{(0)[m'_\pm]}(\rho) \sma C_{\rm L\pm}^{[m'_\pm]}\exp\brkt{\mp m'_\pm\ln\rho}
 = C_{\rm L\pm}^{[m'_\pm]}\rho^{\mp m'_\pm}. 
\eea
Hence, the regularity at the origin requires that
\be
 \pm m_\pm \geq 0, \;\;\;\;\;
 \mp m'_\pm \geq 0.  \label{0:cstrt:2}
\ee
From (\ref{0:cstrt:1}) and (\ref{0:cstrt:2}), the integers~$m_\pm$ and $m'_\pm$ are constrained as 
\bea
 &&0\leq m_+ < q_+n-1, \;\;\;\;\;
 q_-n+1 < m_- \leq 0, \nonumber\\
 &&q_+n+1 < m'_+ \leq 0, \;\;\;\;\;
 0 \leq m'_- < q_-n-1. 
\eea
Thus, the number of zero-modes depends on the charges~$q_\pm$, 
in contrast to the previous case. 
In the absence of Yukawa interactions, it is determined by the charge and 
the flux threading the extra-dimensional space, 
as the index theorem insists~\cite{Atiyah:1968mp,Atiyah:1967ih}. 
However, because the extra-dimensional space is non-compact in our model, 
some of the zero-modes are non-normalizable and are dropped in the spectrum. 
So the number of zero-modes is smaller than that of the compact case. 

When we turn on the Yukawa coupling~$y_1$ or $y_2$, some of the zero-modes allowed in this subsection 
obtain masses via the Yukawa coupling and are decoupled at low energies. 
The number of remaining zero-modes only depends on the vortex number~$n$ and 
is independent of the charges~\cite{Jackiw:1981ee}, as we saw in the previous subsection.


\end{document}